\documentclass[twocolumn,letterpaper,aps,prc,longbibliography,superscriptaddress,nofootinbib,showpacs,floatfix]{revtex4-1}

\usepackage{graphicx}	% Include figure filed
\usepackage{xspace}	% Include xspace

\newcommand{\pt}{\mbox{$p_T$}\xspace}

\newcommand{\raa}{\mbox{$R_{AA}$}\xspace}

\newcommand{\Npart}{\mbox{$N_{\rm part}$}\xspace}
\newcommand{\Ncoll}{\mbox{$N_{\rm coll}$}\xspace}
\newcommand{\gevc}{\mbox{GeV/$c~$}}

\newcommand{\full}{\mbox{$\sqrt{s_{_{NN}}}=$ 200 GeV~}}
\newcommand{\fullb}{\mbox{$\sqrt{s}=$ 200 GeV}}
\newcommand{\pp}{\mbox{$p$$+$$p$~}}

\newcommand{\jpsi}{$J/\psi$}
\newcommand{\jpsib}{$J/\psi~$}

\def\func#1{\left(#1\right)}

\begin{document}


\title{ Measurement of $\Upsilon$(1S+2S+3S) production in p$+$p and 
Au$+$Au collisions at $\sqrt{s_{_{NN}}}=200$ GeV}

\newcommand{\abilene}{Abilene Christian University, Abilene, Texas 79699, USA}
\newcommand{\acadsin}{Institute of Physics, Academia Sinica, Taipei 11529, Taiwan}
\newcommand{\augie}{Department of Physics, Augustana College, Sioux Falls, South Dakota 57197, USA}
\newcommand{\banaras}{Department of Physics, Banaras Hindu University, Varanasi 221005, India}
\newcommand{\barc}{Bhabha Atomic Research Centre, Bombay 400 085, India}
\newcommand{\baruch}{Baruch College, City University of New York, New York, New York, 10010 USA}
\newcommand{\bnlcoll}{Collider-Accelerator Department, Brookhaven National Laboratory, Upton, New York 11973-5000, USA}
\newcommand{\bnlphys}{Physics Department, Brookhaven National Laboratory, Upton, New York 11973-5000, USA}
\newcommand{\caucr}{University of California - Riverside, Riverside, California 92521, USA}
\newcommand{\charlesczech}{Charles University, Ovocn\'{y} trh 5, Praha 1, 116 36, Prague, Czech Republic}
\newcommand{\chonbuk}{Chonbuk National University, Jeonju, 561-756, Korea}
\newcommand{\ciae}{Science and Technology on Nuclear Data Laboratory, China Institute of Atomic Energy, Beijing 102413, P.~R.~China}
\newcommand{\cns}{Center for Nuclear Study, Graduate School of Science, University of Tokyo, 7-3-1 Hongo, Bunkyo, Tokyo 113-0033, Japan}
\newcommand{\colorado}{University of Colorado, Boulder, Colorado 80309, USA}
\newcommand{\columbia}{Columbia University, New York, New York 10027 and Nevis Laboratories, Irvington, New York 10533, USA}
\newcommand{\czechtech}{Czech Technical University, Zikova 4, 166 36 Prague 6, Czech Republic}
\newcommand{\dapnia}{Dapnia, CEA Saclay, F-91191, Gif-sur-Yvette, France}
\newcommand{\debrecen}{Debrecen University, H-4010 Debrecen, Egyetem t{\'e}r 1, Hungary}
\newcommand{\elte}{ELTE, E{\"o}tv{\"o}s Lor{\'a}nd University, H - 1117 Budapest, P{\'a}zm{\'a}ny P. s. 1/A, Hungary}
\newcommand{\ewha}{Ewha Womans University, Seoul 120-750, Korea}
\newcommand{\fit}{Florida Institute of Technology, Melbourne, Florida 32901, USA}
\newcommand{\fsu}{Florida State University, Tallahassee, Florida 32306, USA}
\newcommand{\gsu}{Georgia State University, Atlanta, Georgia 30303, USA}
\newcommand{\hanyang}{Hanyang University, Seoul 133-792, Korea}
\newcommand{\hiroshima}{Hiroshima University, Kagamiyama, Higashi-Hiroshima 739-8526, Japan}
\newcommand{\ihepprot}{IHEP Protvino, State Research Center of Russian Federation, Institute for High Energy Physics, Protvino, 142281, Russia}
\newcommand{\illuiuc}{University of Illinois at Urbana-Champaign, Urbana, Illinois 61801, USA}
\newcommand{\inrras}{Institute for Nuclear Research of the Russian Academy of Sciences, prospekt 60--letiya Oktyabrya 7a, Moscow 117312, Russia}
\newcommand{\instpasczech}{Institute of Physics, Academy of Sciences of the Czech Republic, Na Slovance 2, 182 21 Prague 8, Czech Republic}
\newcommand{\isu}{Iowa State University, Ames, Iowa 50011, USA}
\newcommand{\jaea}{Advanced Science Research Center, Japan Atomic Energy Agency, 2-4 Shirakata Shirane, Tokai-mura, Naka-gun, Ibaraki-ken 319-1195, Japan}
\newcommand{\jinrdubna}{Joint Institute for Nuclear Research, 141980 Dubna, Moscow Region, Russia}
\newcommand{\jyvaskyla}{Helsinki Institute of Physics and University of Jyv{\"a}skyl{\"a}, P.O.Box 35, FI-40014 Jyv{\"a}skyl{\"a}, Finland}
\newcommand{\kek}{KEK, High Energy Accelerator Research Organization, Tsukuba, Ibaraki 305-0801, Japan}
\newcommand{\korea}{Korea University, Seoul, 136-701, Korea}
\newcommand{\kurchatov}{Russian Research Center ``Kurchatov Institute", Moscow, 123098 Russia}
\newcommand{\kyoto}{Kyoto University, Kyoto 606-8502, Japan}
\newcommand{\labllr}{Laboratoire Leprince-Ringuet, Ecole Polytechnique, CNRS-IN2P3, Route de Saclay, F-91128, Palaiseau, France}
\newcommand{\lahorelums}{Physics Department, Lahore University of Management Sciences, Lahore 54792, Pakistan}
\newcommand{\lawllnl}{Lawrence Livermore National Laboratory, Livermore, California 94550, USA}
\newcommand{\losalamos}{Los Alamos National Laboratory, Los Alamos, New Mexico 87545, USA}
\newcommand{\lpc}{LPC, Universit{\'e} Blaise Pascal, CNRS-IN2P3, Clermont-Fd, 63177 Aubiere Cedex, France}
\newcommand{\lund}{Department of Physics, Lund University, Box 118, SE-221 00 Lund, Sweden}
\newcommand{\maryland}{University of Maryland, College Park, Maryland 20742, USA}
\newcommand{\mass}{Department of Physics, University of Massachusetts, Amherst, Massachusetts 01003-9337, USA }
\newcommand{\michigan}{Department of Physics, University of Michigan, Ann Arbor, Michigan 48109-1040, USA}
\newcommand{\muenster}{Institut fur Kernphysik, University of Muenster, D-48149 Muenster, Germany}
\newcommand{\muhlenberg}{Muhlenberg College, Allentown, Pennsylvania 18104-5586, USA}
\newcommand{\myongji}{Myongji University, Yongin, Kyonggido 449-728, Korea}
\newcommand{\nagasaki}{Nagasaki Institute of Applied Science, Nagasaki-shi, Nagasaki 851-0193, Japan}
\newcommand{\natmephi}{National Research Nuclear University, MEPhI, Moscow Engineering Physics Institute, Moscow, 115409, Russia}
\newcommand{\newmex}{University of New Mexico, Albuquerque, New Mexico 87131, USA }
\newcommand{\nmsu}{New Mexico State University, Las Cruces, New Mexico 88003, USA}
\newcommand{\ohio}{Department of Physics and Astronomy, Ohio University, Athens, Ohio 45701, USA}
\newcommand{\ornl}{Oak Ridge National Laboratory, Oak Ridge, Tennessee 37831, USA}
\newcommand{\orsay}{IPN-Orsay, Universite Paris Sud, CNRS-IN2P3, BP1, F-91406, Orsay, France}
\newcommand{\peking}{Peking University, Beijing 100871, P.~R.~China}
\newcommand{\pnpi}{PNPI, Petersburg Nuclear Physics Institute, Gatchina, Leningrad region, 188300, Russia}
\newcommand{\riken}{RIKEN Nishina Center for Accelerator-Based Science, Wako, Saitama 351-0198, Japan}
\newcommand{\rikjrbrc}{RIKEN BNL Research Center, Brookhaven National Laboratory, Upton, New York 11973-5000, USA}
\newcommand{\rikkyo}{Physics Department, Rikkyo University, 3-34-1 Nishi-Ikebukuro, Toshima, Tokyo 171-8501, Japan}
\newcommand{\saispbstu}{Saint Petersburg State Polytechnic University, St. Petersburg, 195251 Russia}
\newcommand{\saopaulo}{Universidade de S{\~a}o Paulo, Instituto de F\'{\i}sica, Caixa Postal 66318, S{\~a}o Paulo CEP05315-970, Brazil}
\newcommand{\seoulnat}{Department of Physics and Astronomy, Seoul National University, Seoul 151-742, Korea}
\newcommand{\stonybrkc}{Chemistry Department, Stony Brook University, SUNY, Stony Brook, New York 11794-3400, USA}
\newcommand{\stonycrkp}{Department of Physics and Astronomy, Stony Brook University, SUNY, Stony Brook, New York 11794-3800,, USA}
\newcommand{\subatech}{SUBATECH (Ecole des Mines de Nantes, CNRS-IN2P3, Universit{\'e} de Nantes) BP 20722 - 44307, Nantes, France}
\newcommand{\tenn}{University of Tennessee, Knoxville, Tennessee 37996, USA}
\newcommand{\titech}{Department of Physics, Tokyo Institute of Technology, Oh-okayama, Meguro, Tokyo 152-8551, Japan}
\newcommand{\tsukuba}{Institute of Physics, University of Tsukuba, Tsukuba, Ibaraki 305, Japan}
\newcommand{\vandy}{Vanderbilt University, Nashville, Tennessee 37235, USA}
\newcommand{\waseda}{Waseda University, Advanced Research Institute for Science and Engineering, 17 Kikui-cho, Shinjuku-ku, Tokyo 162-0044, Japan}
\newcommand{\weizmann}{Weizmann Institute, Rehovot 76100, Israel}
\newcommand{\wigner}{Institute for Particle and Nuclear Physics, Wigner Research Centre for Physics, Hungarian Academy of Sciences (Wigner RCP, RMKI) H-1525 Budapest 114, POBox 49, Budapest, Hungary}
\newcommand{\yonsei}{Yonsei University, IPAP, Seoul 120-749, Korea}
\newcommand{\zagreb}{University of Zagreb, Faculty of Science, Department of Physics, Bijeni\v{c}ka 32, HR-10002 Zagreb, Croatia}
\affiliation{\abilene}
\affiliation{\acadsin}
\affiliation{\augie}
\affiliation{\banaras}
\affiliation{\barc}
\affiliation{\baruch}
\affiliation{\bnlcoll}
\affiliation{\bnlphys}
\affiliation{\caucr}
\affiliation{\charlesczech}
\affiliation{\chonbuk}
\affiliation{\ciae}
\affiliation{\cns}
\affiliation{\colorado}
\affiliation{\columbia}
\affiliation{\czechtech}
\affiliation{\dapnia}
\affiliation{\debrecen}
\affiliation{\elte}
\affiliation{\ewha}
\affiliation{\fit}
\affiliation{\fsu}
\affiliation{\gsu}
\affiliation{\hanyang}
\affiliation{\hiroshima}
\affiliation{\ihepprot}
\affiliation{\illuiuc}
\affiliation{\inrras}
\affiliation{\instpasczech}
\affiliation{\isu}
\affiliation{\jaea}
\affiliation{\jinrdubna}
\affiliation{\jyvaskyla}
\affiliation{\kek}
\affiliation{\korea}
\affiliation{\kurchatov}
\affiliation{\kyoto}
\affiliation{\labllr}
\affiliation{\lahorelums}
\affiliation{\lawllnl}
\affiliation{\losalamos}
\affiliation{\lpc}
\affiliation{\lund}
\affiliation{\maryland}
\affiliation{\mass}
\affiliation{\michigan}
\affiliation{\muenster}
\affiliation{\muhlenberg}
\affiliation{\myongji}
\affiliation{\nagasaki}
\affiliation{\natmephi}
\affiliation{\newmex}
\affiliation{\nmsu}
\affiliation{\ohio}
\affiliation{\ornl}
\affiliation{\orsay}
\affiliation{\peking}
\affiliation{\pnpi}
\affiliation{\riken}
\affiliation{\rikjrbrc}
\affiliation{\rikkyo}
\affiliation{\saispbstu}
\affiliation{\saopaulo}
\affiliation{\seoulnat}
\affiliation{\stonybrkc}
\affiliation{\stonycrkp}
\affiliation{\subatech}
\affiliation{\tenn}
\affiliation{\titech}
\affiliation{\tsukuba}
\affiliation{\vandy}
\affiliation{\waseda}
\affiliation{\weizmann}
\affiliation{\wigner}
\affiliation{\yonsei}
\affiliation{\zagreb}
\author{A.~Adare} \affiliation{\colorado}
\author{S.~Afanasiev} \affiliation{\jinrdubna}
\author{C.~Aidala} \affiliation{\losalamos} \affiliation{\mass} \affiliation{\michigan}
\author{N.N.~Ajitanand} \affiliation{\stonybrkc}
\author{Y.~Akiba} \affiliation{\riken} \affiliation{\rikjrbrc}
\author{R.~Akimoto} \affiliation{\cns}
\author{H.~Al-Bataineh} \affiliation{\nmsu}
\author{H.~Al-Ta'ani} \affiliation{\nmsu}
\author{J.~Alexander} \affiliation{\stonybrkc}
\author{A.~Angerami} \affiliation{\columbia}
% ...
\collaboration{PHENIX Collaboration} \noaffiliation

\date{\today}

\begin{abstract}

%\linenumbers

Measurements of bottomonium production in heavy ion and $p$$+$$p$
collisions at the Relativistic Heavy Ion Collider (RHIC) are
presented. The inclusive yield of the three $\Upsilon$ states,
$\Upsilon(1S+2S+3S)$, was measured in the PHENIX experiment via
electron-positron decay pairs at midrapidity for Au$+$Au and
$p$$+$$p$ collisions at $\sqrt{s_{_{NN}}}=200$ GeV. The 
$\Upsilon(1S+2S+3S)
\rightarrow e^+e^-$ differential cross section at midrapidity was
found to be $B_{\rm ee} d\sigma/dy =$ 108  $\pm$ 38 (stat) $\pm$ 15
(syst) $\pm$ 11 (luminosity) pb in $p$$+$$p$ collisions. The nuclear
modification factor in the 30\% most central Au$+$Au collisions
indicates a suppression of the total $\Upsilon$ state
yield relative to the extrapolation from  $p$$+$$p$ collision data. 
The suppression is consistent with measurements made by STAR at RHIC
and at higher energies by the CMS experiment at the Large Hadron Collider.

\end{abstract}

	
\maketitle

\section{\label{sec:Introduction}Introduction}

One of the main physics programs in relativistic heavy ion collisions
is the study of heavy quarkonia yields, namely charm quark pairs (charmonia)
and bottom quark pairs (bottomonia). At zero temperature, the
binding energy between the heavy quark and anti-quark ($Q\bar{Q}$) in
these vector mesons may be described by an effective potential  consisting
of a confining term at large distance and
Coulomb-like term at short distance~\cite{PhysRevD21_203}.
%All quarkonia states can only be broken if surrounding light quarks transfer an
%energy greater than 1.2 GeV~\cite{Satz:2005hx}.

When the temperature of the medium formed after the collision is
higher than a transition temperature \mbox{$T_c\approx$170 MeV}, the effective
potential between light quark and anti-quark weakens and
deconfines the constituent quarks of mesons and baryons. The Quark-Gluon Plasma (QGP)
formed can be described as a dense, strongly coupled state of
matter which reaches thermalization in less than 1 fm/$c$~\cite{Adcox:2004mh}.

In the QGP medium, the effective color electric potential between $Q$
and $\bar{Q}$ can be screened by the dense surrounding color charges. 
This color screening is similar to the Debye screening observed in
electromagnetic plasmas~\cite{Matsui:1986dk}. The temperature at which
the heavy quark state becomes unbound due to this screening depends on
the corresponding binding energy of the state. Because of the large variation in radii
between the different heavy quarkonia, they are expected to become
unbound at different temperatures.

\begin{figure} [hbt]
    \includegraphics[width=1.0\linewidth]{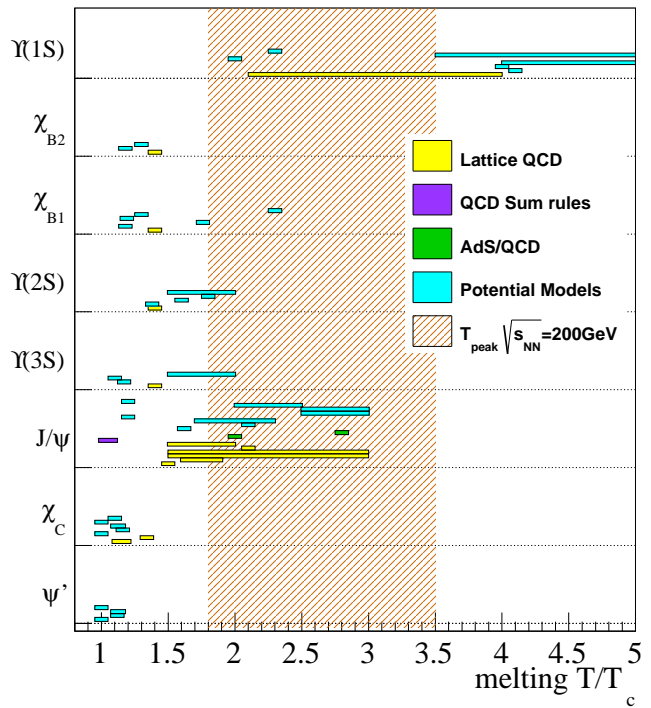}
\caption{\label{fig:qqbar_diccociation_T} (Color online) Compilation
  of medium temperatures relative to the critical temperature
  $\left(T_c\right)$ where quarkonium states are dissociated in the
  quark-gluon plasma. Note that these estimations were performed
  assuming different $T_c$ values. Each horizontal bar corresponds to
  one estimation and its temperature extension (when applied)
  represents the range where the quarkonia state undergoes a mass/size
  modification until it completely melts. Techniques used in
  calculations: Lattice
QCD~\protect\cite{Umeda:2000ym,Asakawa:2003re,Datta:2003ww,Jakovac:2006sf,Aarts:2007pk,Rothkopf:2011db,Aarts:2011sm,Aarts:2010ek,Aarts:2012ka,Aarts:2013kaa,Karsch:2012na},
 QCD sum rules~\protect\cite{Suzuki201328,Morita:2007pt,Morita:2007hv,Song:2008bd,Morita:2009qk,Gubler:2011ua},
  AdS/QCD~\protect\cite{Kim:2007rt,Fujita:2009wc,Noronha:2009da,Grigoryan:2010pj},
  effective field theories~\cite{Brambilla:2008cx,Digal:2005ht} and
  potential
  models~\protect\cite{Alberico:2005xw,Mocsy:2007yj,Mocsy:2007jz,Petreczky:2010tk,Cabrera:2006wh,Riek:2010fk,Riek:2010py,Karsch:2012na}.
  The shaded band from 1.8 to 3.5 $T/T_c$ represents the hydrodynamic
  estimation for the peak temperature reached in Au$+$Au
  collisions at 200 GeV~\protect\cite{PhysRevC81_034911}.}
\end{figure}

There are many theoretical calculations which predict the temperature
at which each quarkonium state is suppressed by color screening. A
compilation of results can be found in~\cite{Suzuki201328}, including
lattice quantum chromodynamics 
(QCD)~\cite{Umeda:2000ym,Asakawa:2003re,Datta:2003ww,Jakovac:2006sf,Aarts:2007pk,Rothkopf:2011db,Aarts:2011sm,Aarts:2010ek,Aarts:2012ka,Aarts:2013kaa,Karsch:2012na},
QCD sum
rules~\cite{Suzuki201328,Morita:2007pt,Morita:2007hv,Song:2008bd,Morita:2009qk,Gubler:2011ua},
AdS/QCD~\cite{Kim:2007rt,Fujita:2009wc,Noronha:2009da,Grigoryan:2010pj},
resummed perturbation theory~\cite{Laine:2007gj,Wong:2004zr},
effective field theories~\cite{Brambilla:2008cx,Digal:2005ht}, and
potential
models~\cite{Alberico:2005xw,Mocsy:2007yj,Mocsy:2007jz,Petreczky:2010tk,Cabrera:2006wh,Riek:2010fk,Riek:2010py,Karsch:2012na}.
Figure~\ref{fig:qqbar_diccociation_T} shows the dissociation
temperature range for several quarkonium states as expected from these
models. Besides the different techniques used in these calculations,
the melting range also depends on the choice of the transition
temperature, the use of the internal energy or the free energy of the
system for the temperature dependence of the heavy quark potential and
the criteria adopted for defining the dissociation point.  No cold
nuclear matter effects have been considered in these estimations.

%%%%%%%%%%%%%%%%%%%%%%%%%%%%%%%%%%%%%%%%%%%%%%%%%%%%%%%%% Fig_1

A comparison between hydrodynamical model calculations and the PHENIX 
thermal photon data~\cite{PhysRevC81_034911} suggests that the peak 
temperature of the medium formed at RHIC in central Au$+$Au collisions at 
\full lies in the region between 300 and 600 MeV, or 1.8 $T_c$ and 3.5 
$T_c$. The majority of the estimates shown in 
Fig.~\ref{fig:qqbar_diccociation_T} indicates that only the ground states, 
the \jpsib and $\Upsilon$(1S), remain bound at these temperatures.

PHENIX reported a strong suppression of the \jpsib yield in central
Au$+$Au collisions compared to binary collision scaling from \pp
yields~\cite{PhysRevLett98_232301,PhysRevC84_054912}. According to
measurements performed in \pp collisions at RHIC, \mbox{(42 $\pm$
  9)\%} of the \jpsib yield comes from $\chi_c$ and $\psi^{\prime}$
decays~\cite{Adare:2011vq}. The complete suppression of these
states in Au$+$Au collisions can explain only part of the suppression
seen for the \jpsi. There are other possible contributions to $J/\psi$
suppression and therefore the interpretation of the data is not
straightforward.  Other mechanisms of suppression include initial and
final state cold nuclear matter effects, studied in $d$$+$Au collisions
by PHENIX~\cite{PhysRevLett107_142301,PhysRevC87_034904}. 
There are also effects that can reduce the suppression.
The dissociated charm (and anti-charm) quark can undergo multiple
scatterings and recombine with its former partner, once the medium
cools down. In addition, the presence of about 6-20 open charm pairs in each
central Au$+$Au collision at RHIC \footnote{This estimation is based on
the $c-\bar{c}$ total cross section reported
in~\cite{PhysRevLett97_252002} and 1000 binary collisions in very
central Au$+$Au events.}, provides a good chance that the ground state
charmonium was formed by coalescence of uncorrelated charm and
anti-charm quarks present in the medium~\cite{PhysRevC63_054905}.
Thus, even if all the initially produced $J/\psi$s are dissociated in
the QGP medium, $J/\psi$s can be re-created at a later stage by the
coalescence process.

The probability for creating a bottomonium state through coalescence
is quite small at \full, given that only about 0.07 $b\bar{b}$ 
pairs per
central event are produced \footnote{Estimation based on the total
$b\bar{b}$ cross section published in~\cite{Adare:2009ic}.}.
Therefore, bottomonium states are a better
probe of color screening in Au$+$Au collisions at RHIC.
Figure~\ref{fig:qqbar_diccociation_T} shows that no lattice QCD or
potential model calculation predicts that $\Upsilon$(1S) will melt at
a temperature lower than around 2 $T_c$. This is an outcome of the
tighter binding energy and smaller radius of the 1S state compared to
other quarkonium states.  Some calculations suggest the ground state
charmonium is dissociated at a temperature close to
$T_c$~\cite{Gubler:2011ua,Mocsy:2007jz,Riek:2010fk}.

Bottomonia have been measured mostly in the dilepton channel with a
branching ratio around
2.5\%~\cite{Beringer:1900zz}. Table~\ref{tab:upsilon_frac} lists the
fraction of the three $\Upsilon$ states present in the dilepton
spectrum as measured at Fermilab and the Large Hadron Collider (LHC) 
by E866/NuSea~\cite{PhysRevLett100_062301},
CDF~\cite{PhysRevLett75_4358}, LHCb~\cite{LHCb:2012aa} and
CMS~\cite{Khachatryan:2010zg}.  No significant variations on the
relative yields have been observed in spite of the broad collision
energy range of these experiments or whether the anti-proton was one
of the collision particles or not. The ground state 
$\Upsilon$(1S) has many feed-down contributions from excited
states. The CDF experiment reported the fraction of these
contributions~\cite{PhysRevLett84_2094}, which can be seen
in Table~\ref{tab:upsilon1s_source}.

%%%%%%%%%%%%%%%%%%%%%%%%%%%%%%%%%%%%%%%%%%%%%%%%%%%%% Table_I
\begin{table}
  \caption{\label{tab:upsilon_frac} Composition of the $\Upsilon$
    family in the dilepton channel as measured by
    E866/NuSea~\cite{PhysRevLett100_062301},
    CDF~\cite{PhysRevLett75_4358}, LHCb~\cite{LHCb:2012aa} and
    CMS~\cite{Khachatryan:2010zg}. Fractions are in \% and only
    statistical uncertainties are shown.}
  \begin{ruledtabular} \begin{tabular}{lcccc}
    Exp. & system & $\Upsilon$(1S) & $\Upsilon$(2S) &
    $\Upsilon$(3S) \\
    & & {\footnotesize 9.46 $\frac{GeV}{c^2}$} &  {\footnotesize 10.02
      $\frac{GeV}{c^2}$}
    &  {\footnotesize 10.36 $\frac{GeV}{c^2}$} \\\hline
    E866 & \pp $\sqrt{s}=$39 GeV &
    69.1 $\pm$ 1.0 & 22.2 $\pm$ 0.9 & 8.8 $\pm$ 0.6 \\
    CDF & $p+\bar{p}$ $\sqrt{s}=$1.8 TeV &
    72.6 $\pm$ 2.8 & 17.6 $\pm$ 1.7 & 9.7 $\pm$ 1.4 \\
    LHCb & \pp $\sqrt{s}=$7 TeV &
    73.0 $\pm$ 0.3 & 17.9 $\pm$ 0.2 & 9.0 $\pm$ 0.2\\
    CMS & \pp $\sqrt{s}=$7 TeV &
    71.6 $\pm$ 1.3 & 18.5 $\pm$ 0.8 & 10.0 $\pm$ 1.3\\
  \end{tabular} \end{ruledtabular}
\end{table}

%%%%%%%%%%%%%%%%%%%%%%%%%%%%%%%%%%%%%%%%%%%%%%%%%%%%% Table_II
\begin{table}
  \caption{\label{tab:upsilon1s_source} Feed-down fractions of the
    $\Upsilon$(1S) state in \pp collisions as measured by CDF for $p_T>8$ \gevc
    \cite{PhysRevLett84_2094}.}
  \begin{ruledtabular} \begin{tabular}{lc}
    Source & fraction $\pm$ stat $\pm$ syst \\\hline
    Direct $\Upsilon$(1S) & 0.509 $\pm$ 0.082 $\pm$ 0.090 \\
    $\Upsilon$(2S)           & 0.107 $\pm$ 0.077 $\pm$ 0.048 \\
    $\Upsilon$(3S)           & 0.008 $\pm$ 0.006 $\pm$ 0.004 \\
    $\chi_{\rm B1}$                  & 0.271 $\pm$ 0.069 $\pm$ 0.044 \\
    $\chi_{\rm B2}$                  & 0.105 $\pm$ 0.044 $\pm$ 0.014 \\    
  \end{tabular} \end{ruledtabular}
\end{table}

Fermilab experiments found no modification of the relative yields in cold 
nuclear matter as measured in $p$+$d$ \cite{PhysRevLett100_062301} and 
$p$+A \cite{PhysRevLett66_2285}. The initial state effects on bottomonia 
production were investigated by E605~\cite{PhysRevD43_2815}, 
E772~\cite{PhysRevLett66_2285} and 
E866/NuSea~\cite{PhysRevLett100_062301} in $p$+A collisions at 
$\sqrt{s_{_{NN}}}$=38.8 GeV with targets of $^2$H, C, Ca and Fe. The 
$\Upsilon$ yields are suppressed by $\sim$5\% for incident gluon momentum 
fraction $x_2\sim 0.1$. The suppression gets stronger for larger $x_2$, 
reaching a level of $\sim$15\% at $x_2\sim$0.3. PHENIX measured the medium 
modification of the $\Upsilon$ family (1S+2S+3S) yield in $d$$+$Au 
collisions at \full~\cite{PhysRevC87_044909}. The result is consistent 
with no modification within the large statistical uncertainties at 
$x_2\sim 10^{-2}$ and presents an one standard-deviation suppression at 
$x_2\sim 0.2$, which is consistent with the Fermilab results and the
STAR experiment at midrapidity in $d$$+$Au collisions \cite{tagkey2014127}.  The RHIC 
results can be accounted for by a combination of initial state effects, 
calculated by the parton modification function 
EPS09~\cite{Aarts:2011sm}, and quarkonium breakup when crossing 
the cold nuclear matter.

QGP effects on $\Upsilon$ production were studied at the
LHC by the CMS experiment~\cite{Chatrchyan:2012lxa} using Pb+Pb
collisions at $\sqrt{s_{_{NN}}}$=2.76~TeV.
The excited state $\Upsilon$(2S) is more suppressed than the
$\Upsilon$(1S) and the $\Upsilon$(3S) state is not seen in CMS data. This is
qualitatively consistent with expectations of the effects of color
screening from several models discussed earlier. The question which 
arises is whether or not the suppression also happen at lower energies
and in an environment with a much smaller number of bottom quarks
present in the medium.

This paper reports the measurement of the inclusive $\Upsilon$
(1S+2S+3S) yield at $|y|<0.35$ in Au$+$Au collisions at \fullb. Section II
describes the experimental apparatus and the data sample used in the
measurement. Section III details the signal extraction, detector
response and systematic uncertainties involved in this measurement. The
results and comparisons with other measurements and models are
presented in Section IV. The final conclusions are presented in
Section V.

\section{\label{sec:exp and data}Experimental Apparatus and Data Set}

%%%%%%%%%%%%%%%%%%%%%%%%%%%%%%%%%%%%%%%%%%%%%%%%%%%%%%%%% Fig_2
\begin{figure}[tbh]
    \includegraphics[width=1.0\linewidth]{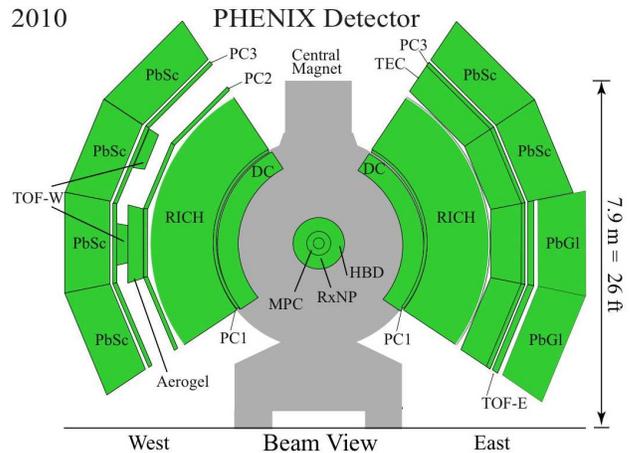}
    \caption{\label{fig:PHENIX}(Color online) 
The PHENIX Central Arm Spectrometers for the 2010 data taking period.
    }
\end{figure}

The PHENIX experiment measures quarkonia at midrapidity through their 
dielectron decays with the two-arm central 
spectrometers~\cite{Adcox:2003zp} shown in Fig.~\ref{fig:PHENIX}.  The 
central arm detectors measure electrons, photons, and hadrons over 
pseudorapidity of $\vert \eta \vert$ $<$ 0.35 with each arm covering 
azimuthal angle $\Delta \phi = \pi$/2. Charged particle tracks in the 
central arms are reconstructed using the drift chambers (DC), the pad 
chambers, and the collision point. Electron candidates are selected 
using information from the ring-imaging \v{C}erenkov detector (RICH) and 
the electromagnetic calorimeter (EMCal)~\cite{Adare:2011vq}. The total 
radiation length before the DC during the 2006 \pp run was 
0.4\%. During the 2010 Au$+$Au run more material was introduced from the 
hadron blind detector (HBD) which added 2.4\% radiation lengths to what 
the detector had in 2006.  In the 2010 run, the magnetic field 
configuration was also modified to cancel the field in the HBD volume, 
decreasing the momentum resolution by about 25\%.

Beam interactions were selected with a minimum bias (MB) trigger that
requires at least one hit (two in Au$+$Au collisions) per beam crossing in each of the two
beam-beam counters (BBC) placed at 3.0 $<$ $\vert \eta \vert$ $<$
3.9. In the Au$+$Au data set, this was the only trigger used. A dedicated EMCal-RICH-Trigger (ERT) was used in
coincidence with the MB trigger during the 2006 \pp data acquisition. The ERT
required a minimum energy in any 2$\times$2 group of EMCal towers,
corresponding to $\Delta \eta \times \Delta \phi \approx 0.02 \times 0.02$
rad., plus associated hits in the RICH. The minimum EMCal energy
requirement was 400 MeV for the first half of the run and 600 MeV for
the second half. 

The collision point along the beam direction was determined with a
resolution of 1.5 cm in $p$$+$$p$ collisions and 0.5 cm in Au$+$Au
collisions, by using the difference between the time
signals measured between the two BBC detectors. The collision
point was required to be within $\pm$30 cm of the nominal center of
the detector in \pp collisions and $\pm$20 cm in Au$+$Au collisions. The
2006 data sample was taken from N$_{\rm pp}$ = 143 billion 
minimum bias events, corresponding to an integrated luminosity of 6.2
pb$^{-1}$. The 2010 data sample was obtained from \mbox{N$_{\rm 
  AuAu}$ = 5.41 billion} minimum bias events, corresponding to 0.9
nb$^{-1}$.

\begin{figure*}[tbh]
    \includegraphics[width=0.49\linewidth]{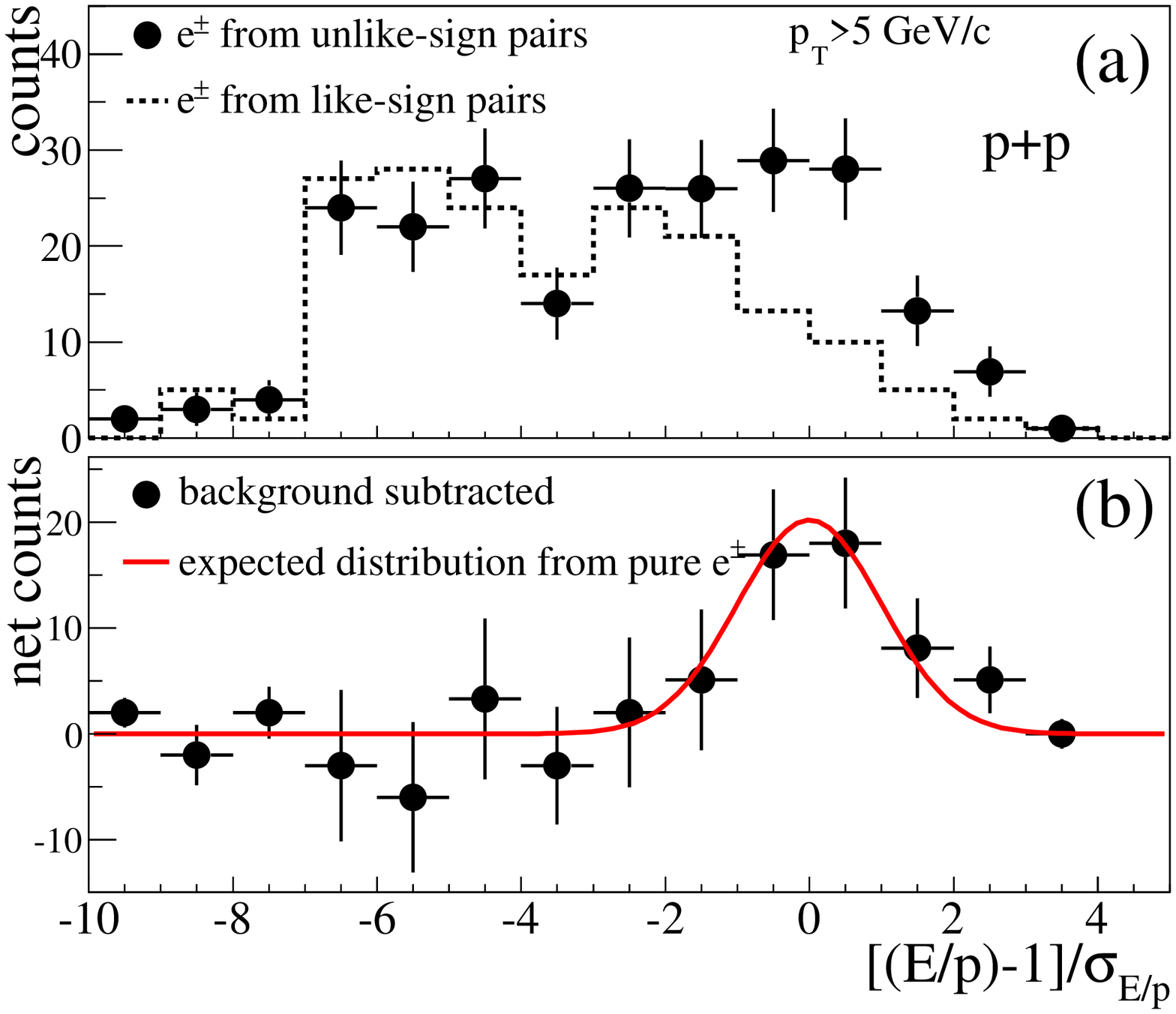}
    \includegraphics[width=0.49\linewidth]{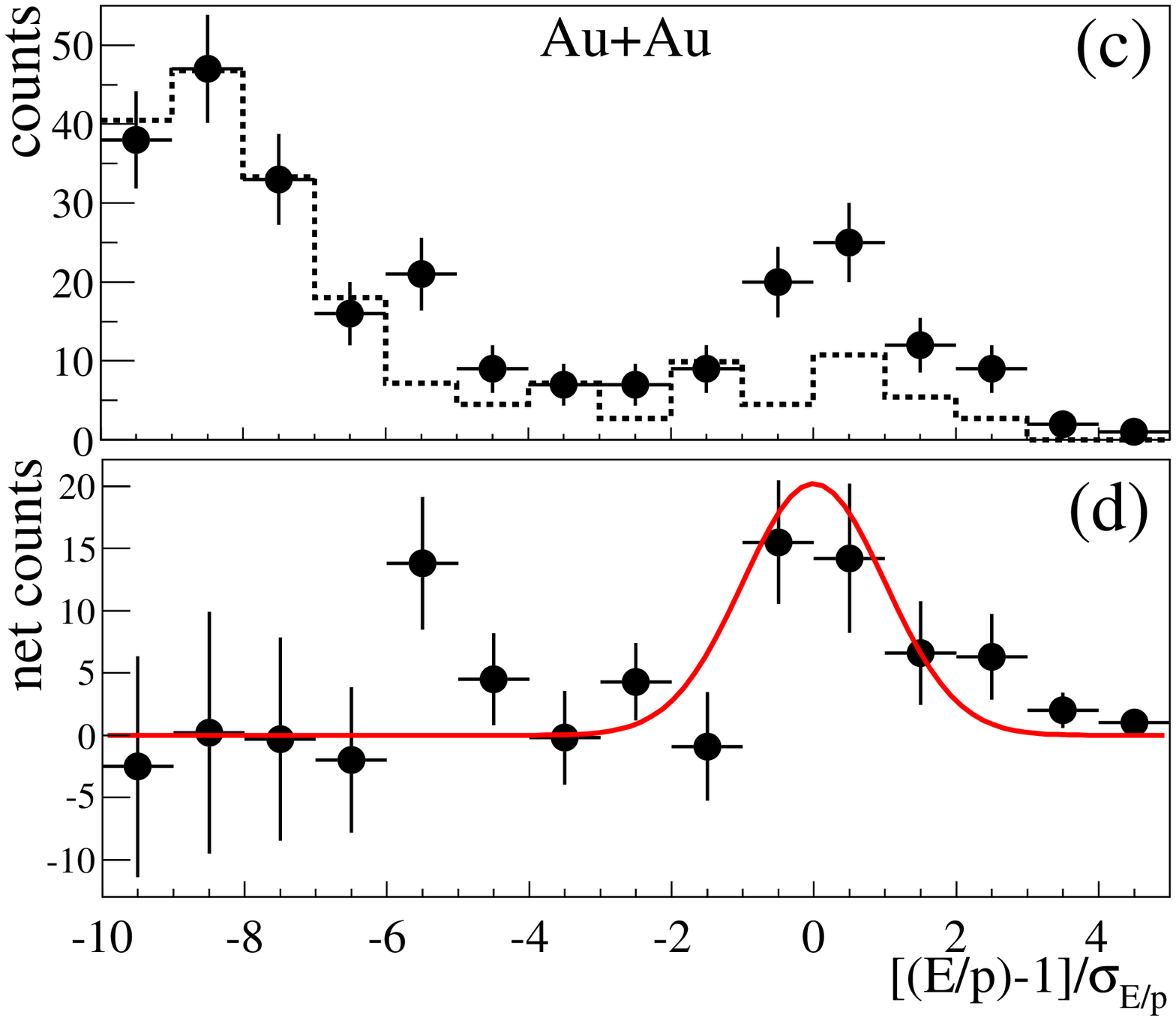}
    \caption{\label{fig:dep} (Color online) Distribution of the parameter used to
      identify electrons with the EMCal. $E/p$ is the ratio between the energy
      deposited by the particle in the EMCal cluster and its momentum,
      $\sigma_{\rm E/p}$ is the variance of the expected
      energy/momentum expected for electrons. The sample shown in (a) from \pp 
      collisions and (c) from Au$+$Au collisions is from unlike-sign electron pairs
      (containing signal+combinatorial background) and like-sign 
    pairs (containing only background). (b) and (d) are the background
    subtracted distributions along with the expected line shape from pure
    electrons.} 
\end{figure*}

In \pp collisions, electron candidates were identified by requiring at
least one fired phototube within an annulus 3.4 $< R_{\rm ring}[
cm] <$ 8.4 centered in the projected track position on the
RICH. The RICH is filled with a CO$_2$ radiator at 1 atm.  Pions with
momentum larger than 4.8 \gevc can also produce 
\v{C}erenkov light in the RICH.  Electron candidates are also
required to be associated with an energy cluster in the EMCal that
falls within 4$\sigma_{\rm  position}$ of the projected track position
and within 4$\sigma_{\rm E/p}$ of the expected energy/momentum ratio
for electrons, where $\sigma$ represents one standard deviation in the
position and energy+momentum resolution of the EMCal+DC determined using
electrons from fully reconstructed Dalitz decays.
Figure~\ref{fig:dep} shows the distribution of the parameter
used to select electrons in the EMCal using electron candidates 
used in high-mass dielectrons with $p_T>$5 \gevc, above the
\v{C}erenkov threshold. Hadron contamination appears as an enhancement of this
distribution for negative values. The distribution,
after subtracting the background mainly composed of hadrons,
represents a clean sample of electrons for $(E/p)-1<4\sigma_{E/p}$.

%%%%%%%%%%%%%%%%%%%%%%%%%%%%%%%%%%%%%%%%%%%%%%%%%%%%%%%%% Fig_3

In the Au$+$Au analysis, the cuts were optimized by looking at the 
parameters in the detector simulations using generated
$\Upsilon\rightarrow e^+e^-$ decays embedded into real data for the
signal, and the real data like-sign dielectrons as a  
background. As a result of the optimization, we require:

\begin{itemize}
\item at least two fired phototubes within an annulus 
\mbox{$3.4<R_{\rm ring}[{\rm cm}]<8.4$} centered in the projected track 
position on the RICH
\item $\chi^2/npe0<25$, a variable defined as $\chi^2$-like shape of
the RICH ring associated to the track over the number of
photoelectrons detected in the ring
\item the displacement between the ring centroid and the track
projection should be smaller than 7cm
\item EMCal cluster-track matching should be smaller 
than 3$\sigma_{\rm  position}$
\item EMCal cluster energy/momentum  ratio should be larger 
than -2.5$\sigma_{\rm E/p}$.
\end{itemize}

These tighter cuts allowed a better hadron rejection as can be seen in
Figure \ref{fig:dep}-c compared to the \pp sample in Figure \ref{fig:dep}-a.

%%%%%%%%%%%%%%%%%%%%%%%%%%%%%%%%%%%%%%%%%%%%%%%%%%%%%%%%% Fig_4
\begin{figure}[htb]
	\includegraphics[width=1.0\linewidth]{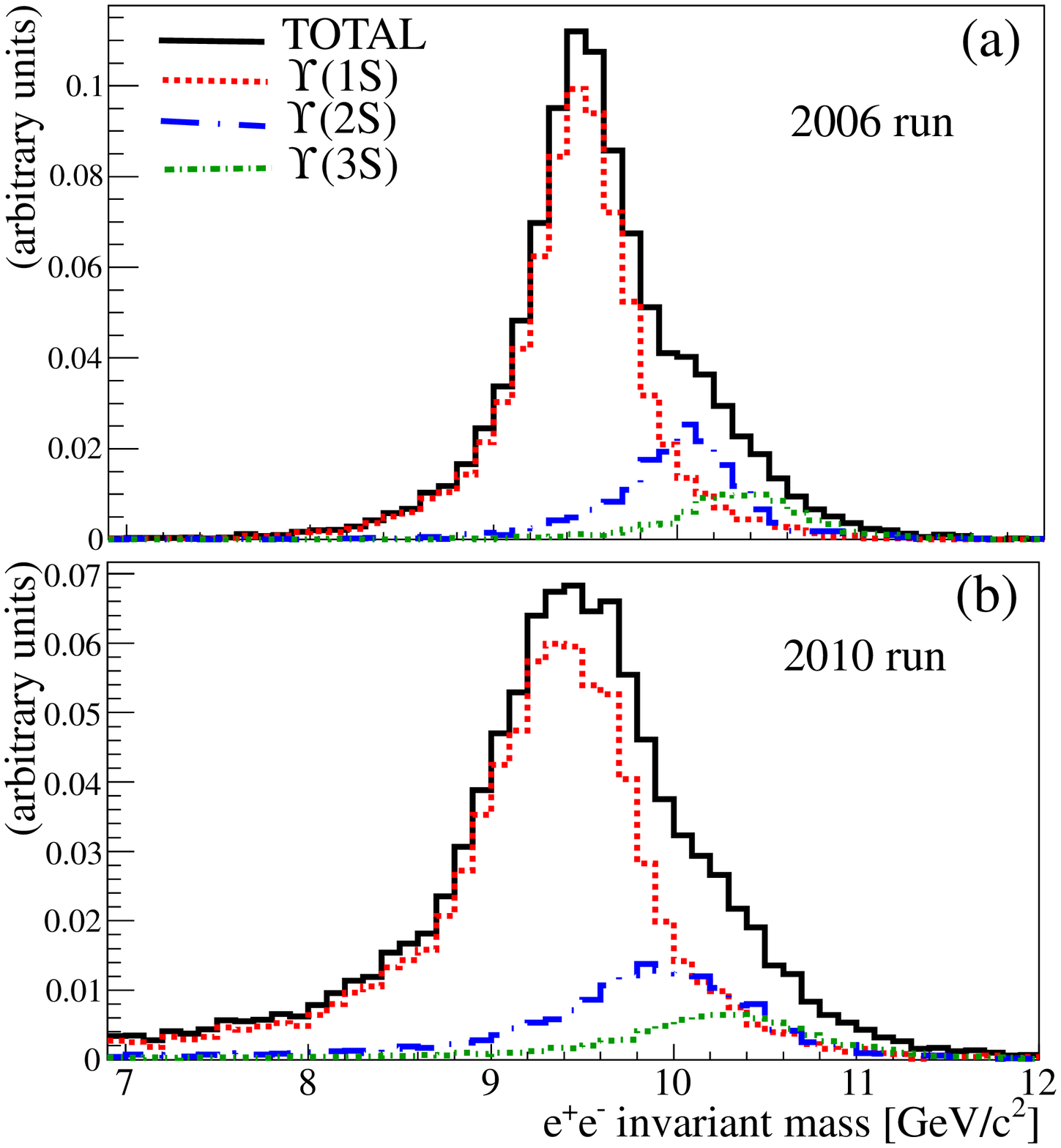}
	\caption{\label{fig:upsilon_1s2s3s}(Color online) Invariant mass distribution
          of simulated $\Upsilon$ (1S+2S+3S) using the PHENIX detector
          simulation and relative $\Upsilon$ yields from CDF
          experiment~\cite{PhysRevLett75_4358} in 2006 run (a) and 2010
          run (b) detector configurations.}
\end{figure}

Figure~\ref{fig:upsilon_1s2s3s} shows the reconstructed invariant mass
distribution for the three $\Upsilon$ states from PHENIX detector
simulations in the 2006 \pp run configuration and in 2010 Au$+$Au
configuration. The detector is not able to separate the three states and a
single peak should be observed. In the 2010 detector configuration the
addition of more material in the detector introduced more
bremsstrahlung for the electrons increasing the low mass tail of the peaks.

\section{Analysis Procedure}
\label{sec:analysis}

\subsection{Dielectrons from $\Upsilon$ in the Central Arms}

\begin{figure}[tbh]
	\includegraphics[width=1.0\linewidth]{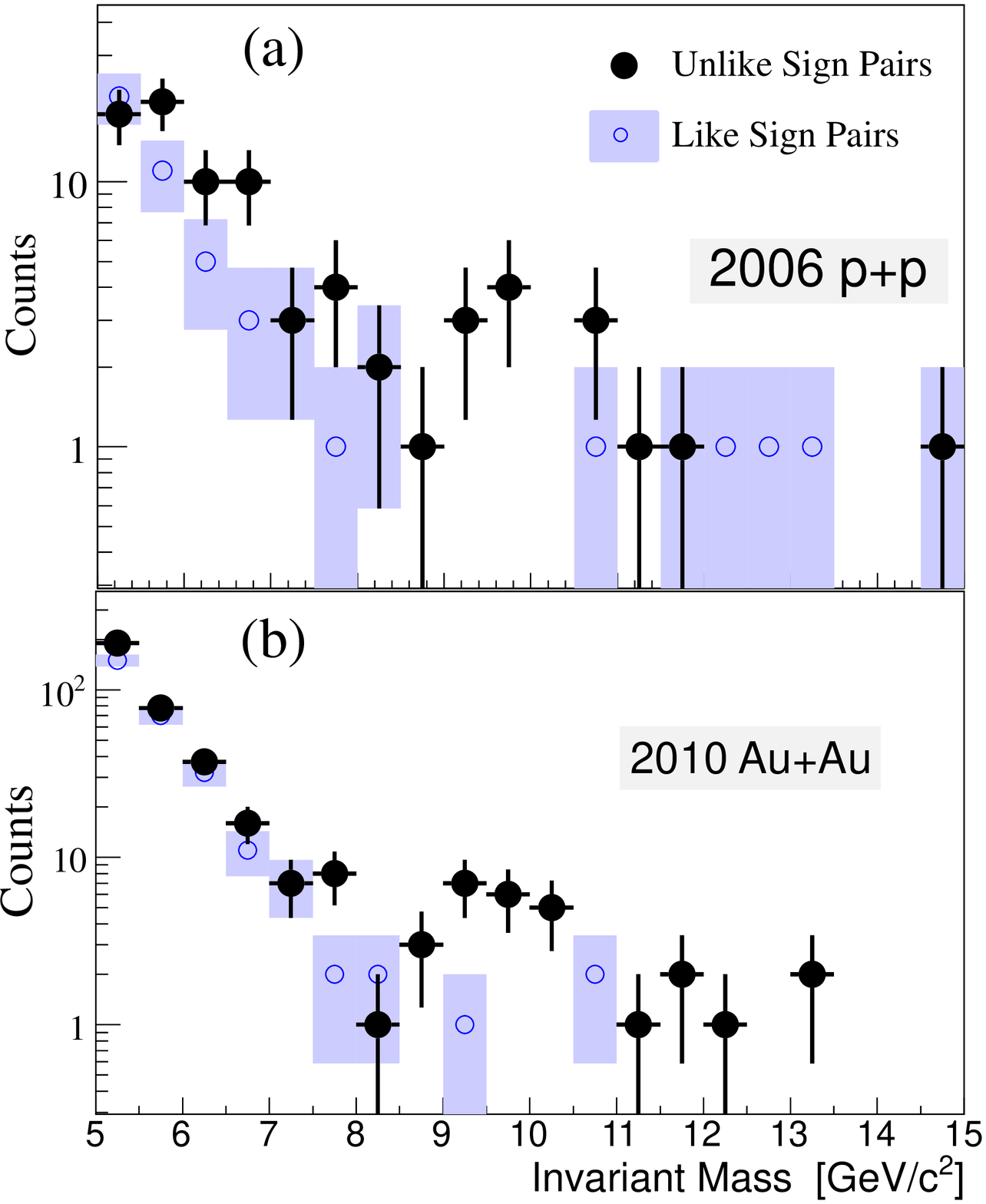}
	\caption{\label{fig:mass_spectra}(Color online) Invariant mass distribution
          of unlike-sign and like-sign dielectrons in the $\Upsilon$
          mass region taken from \pp (a), and Au$+$Au collisions (b).}
\end{figure}

The invariant mass was calculated for all electron pairs. Dielectron
contributions to $\Upsilon$ decays are clearly identified as a peak in
the unlike-sign invariant mass distributions around the $\Upsilon$ 
mass range \mbox{$8.5 < M_{\rm ee}$[GeV/$c^2$]$ < 11.5$} 
(Fig.~\ref{fig:mass_spectra}). There
were 12 unlike and one like-sign dielectron within this mass region
from the \pp sample. In the Au$+$Au sample there were 22 unlike and 3
like-sign pairs in the same mass region.

%%%%%%%%%%%%%%%%%%%%%%%%%%%%%%%%%%%%%%%%%%%%%%%%%%%%%%%%% Fig_5

Figure~\ref{fig:invmass_fit_run6pp} shows the \pp dielectron mass
spectrum over an extended mass region after the like-sign distribution
(used to estimate combinatorial background) has been subtracted from
the unlike-sign data. 
Figure~\ref{fig:corr_fits} shows the same invariant mass
spectrum in the $\Upsilon$ mass region for \pp and Au$+$Au data. 
The line shape of the $\Upsilon$ mass peak determined from
simulations (Fig.~\ref{fig:upsilon_1s2s3s}) cannot be validated by
the real data given the low statistics in both \pp and Au$+$Au
samples. In addition, the relative contributions from different
$\Upsilon$ states are unknown in Au$+$Au data. The
number of $\Upsilon$ counts was determined from a direct count of
unlike-sign and like-sign dielectrons in the $\Upsilon$ mass region
and the fraction of correlated background $f_{\rm cont}$ in the same
mass range. Given the low counts for the signal and background, Poisson
statistics precludes the use of a simple subtraction. Therefore, the
$\Upsilon$ signal is determined from

\begin{equation}
  \label{eq:upsilon_counts}
  N_{\rm \Upsilon} = <s>_P(1-f_{\rm cont}),
\end{equation}

\noindent where $<s>_P$ is the average signal from a joint Poisson distribution
from the foreground unlike-sign $f$ and background like-sign 
$b$ dielectron counts in the $\Upsilon$ mass region~\cite{Adare:2011vq}

\begin{eqnarray}
  \label{eq:tannembaum_prob}
  P(s) &=&
  \sum_{k=0}^{f}\frac{(b+f-k)!}{b!(f-k)!}\frac{1}{2}\left(\frac{1}{2}
  \right)^{b+f-k}\frac{s^ke^{-s}}{k!},
\end{eqnarray}

\noindent and the statistical uncertainty corresponds to one standard deviation of the $P(s)$ distribution.

\subsection{Estimation of the continuum contribution}
\label{sec:continuum}

The correlated background underneath the $\Upsilon$ region is
determined from fits of the expected mass dependence of Drell-Yan,
correlated electrons from $B$ meson decays and possible contamination
of hadrons within jets.  

The Drell-Yan contribution was estimated from next-to-leading order
(NLO) QCD calculations~\cite{Kubar:1980zv}.  These calculations are
known to reproduce lower and higher energy data at
Fermilab~\cite{Webb:2003ps,PhysRevD63_011101}. The calculated cross
section was used to generate dielectrons propagated through the
{\sc geant}~\cite{GEANT} based detector simulation. The Drell-Yan
contribution is modified by isospin and initial state effects in Au$+$Au
collisions. After calculating the Drell-Yan cross section for $p$+$n$
and $n$+$n$ collisions, we found that the Au$+$Au cross section per
binary collision is $f_{\rm iso}=$89\% of that of \pp collisions
because of the isospin effect. The
initial state effects were accounted for by using a parton modification
factor from the EPS09 parametrization,
$R_q^{DY}\func{Q^2,x_1,x_2}$, for both Au nuclei.
The expected Drell-Yan yield in Au$+$Au collisions
$\left(Y_{DY}^{\rm AuAu}\right)$ relative to the yield in $p$$+$$p$ 
collisions $\left(Y_{DY}^{pp}\right)$ is:

\begin{equation}
  \label{eq:DY}
  \frac{Y_{\rm DY}^{\rm AuAu}\func{M_{\rm ee}}}{N_{\rm coll}} =  Y_{\rm DY}^{pp}
  \left(M_{\rm ee}\right) \cdot f_{\rm iso} \cdot R_q^{DY}\func{Q^2,x_1,x_2},
\end{equation}

\noindent where $N_{\rm coll}$ is the number of binary
collisions. $Q^2$, $x_1$ and $x_2$ are taken event-by-event from a
{\sc pythia} simulation~\cite{Sjostrand:2006za}. Theoretical uncertainties
from the NLO calculation, EPS09 quark modification factor
$\func{R_q^{DY}\func{Q^2,x_1,x_2}}$ and overall detector response were accounted
for in the Drell-Yan contribution.

%%%%%%%%%%%%%%%%%%%%%%%%%%%%%%%%%%%%%%%%%%%%%%%%%%%%%%%%% Fig_6
\begin{figure}[htb]
  \includegraphics[width=1.0\linewidth]{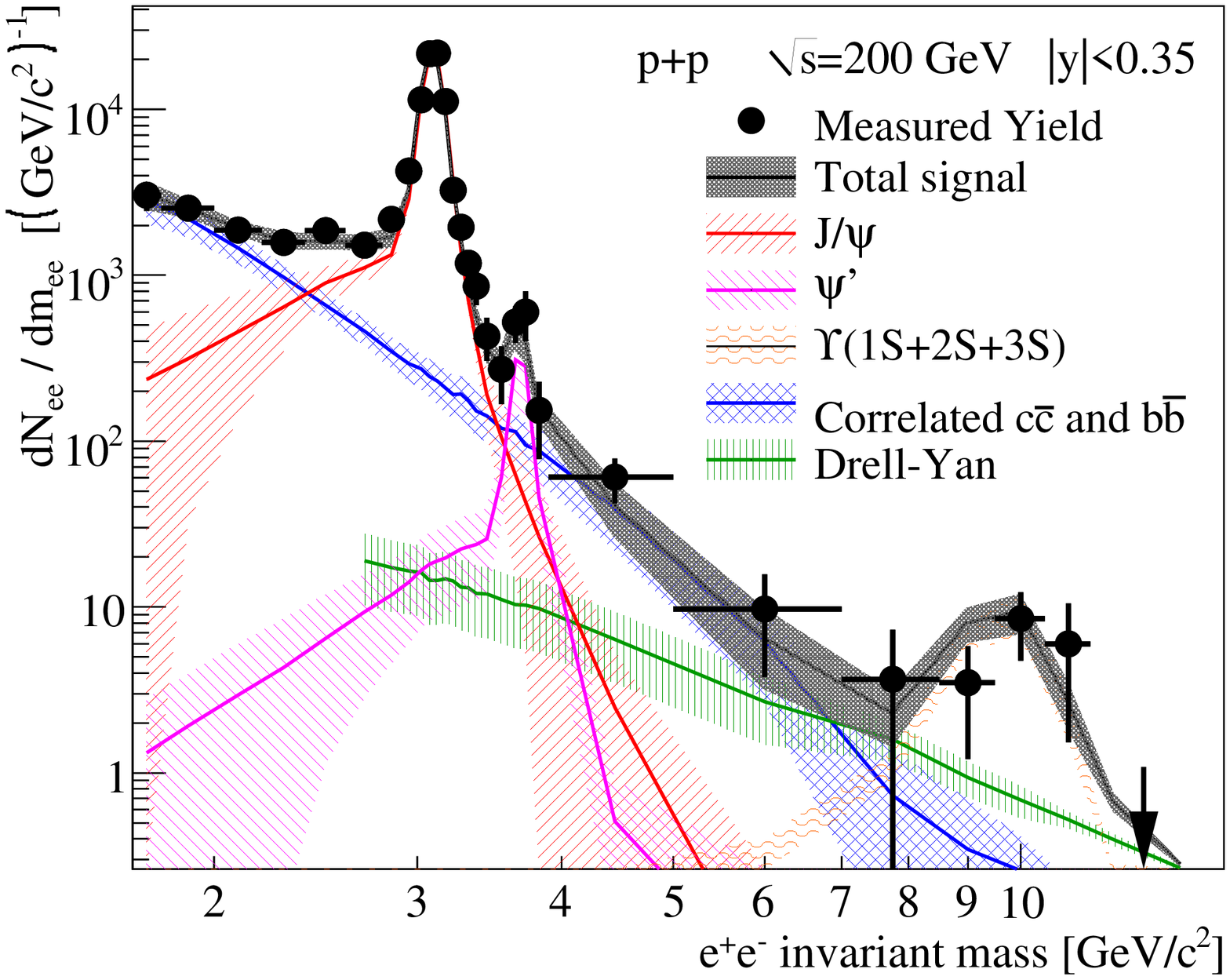}
  \caption{\label{fig:invmass_fit_run6pp}(Color online) Fitted components to the
    correlated dielectron mass spectrum in the \pp sample. The bands
    correspond to the uncertainties obtained from the fit, changes in
    the heavy flavor generator and theoretical uncertainties in the
    Drell-Yan contribution.}
\end{figure}

The line shape of the correlated high-mass dielectron distribution
from heavy flavor decays in \pp collisions was studied in detail
in~\cite{Adare:2011vq}. Two approaches were used: (1) a dielectron
generator using the measured $p_T$ distribution of single electrons from
heavy flavor with a random opening angle and (2) a heavy flavor
simulation from {\sc pythia} in the hard scattering mode to emulate
NLO contributions. Both generated dielectron distributions were introduced
into the detector simulation and reconstructed like the real data. The
mass distribution from heavy flavor decays was normalized 
according to a fit to the dielectron spectrum starting at an
invariant mass at 1.7 GeV/$c^2$, thus including the \jpsib and the
$\psi^{\prime}$ peaks. Figure~\ref{fig:invmass_fit_run6pp} shows the
overall dielectron fit extended to the $\Upsilon$ region. The
uncertainty bands represent the quadratic sum of the fit uncertainties
and the differences between the approaches (1) and (2). The Drell-Yan
band represents the quadratic sum of theoretical uncertainties and
detector response uncertainties. The extrapolation of the heavy flavor
contribution to the $\Upsilon$ mass range \mbox{$8.5 < M_{\rm
    ee}$[GeV/$c^2$]$ < 11.5$} in \pp data yields 0.29 $\pm$ 0.12 counts,
which corresponds to \mbox{3.9 $\pm$ 1.7 pb}. The {\sc pythia} simulation,
including parton shower terms, yields an estimate that the correlated bottom
contribution in this mass range is 3.2 pb, in agreement with the fit
extrapolated result.

Jets can contribute to the correlated background in two ways: Dalitz 
decays from $\pi^0$ pairs within the jet and correlated hadron pair 
contamination.  For a $\pi^0$ pair to produce a correlated electron pair 
in the $\Upsilon$ mass region, each of the $\pi^0$s should have a 
transverse momentum larger than the mass of the $\Upsilon$, which is a 
possibility ruled out by the current statistics.  Figure \ref{fig:dep} 
shows a not significant hadron contamination in the high-mass 
dielectrons in \pp 
data after combinatorial background subtraction. Hadron contamination was 
found to be negligible within uncertainties. Contributions from 
electron-hadron 
correlations are also assumed to be negligible.

The resulting continuum fraction in the selected mass range is
$f_{\rm cont}^{pp}=13 \pm 4$\% in the \pp sample. The continuum fraction
was also determined with a maximum likelihood fit using the combinatorial background,
Drell-Yan, $B$ meson and $\Upsilon$ line shapes with free parameters for their
scales, except the combinatorial background which has a fixed scale. The
total continuum found in this manner was consistent with that
estimated with a fixed Drell-Yan scale. The fit (without any hadron
contribution) provides a good description of the mass distribution.

%%%%%%%%%%%%%%%%%%%%%%%%%%%%%%%%%%%%%%%%%%%%%%%%%%%%%%%%% Fig_7
\begin{figure*}[t]
	\includegraphics[width=0.47\linewidth]{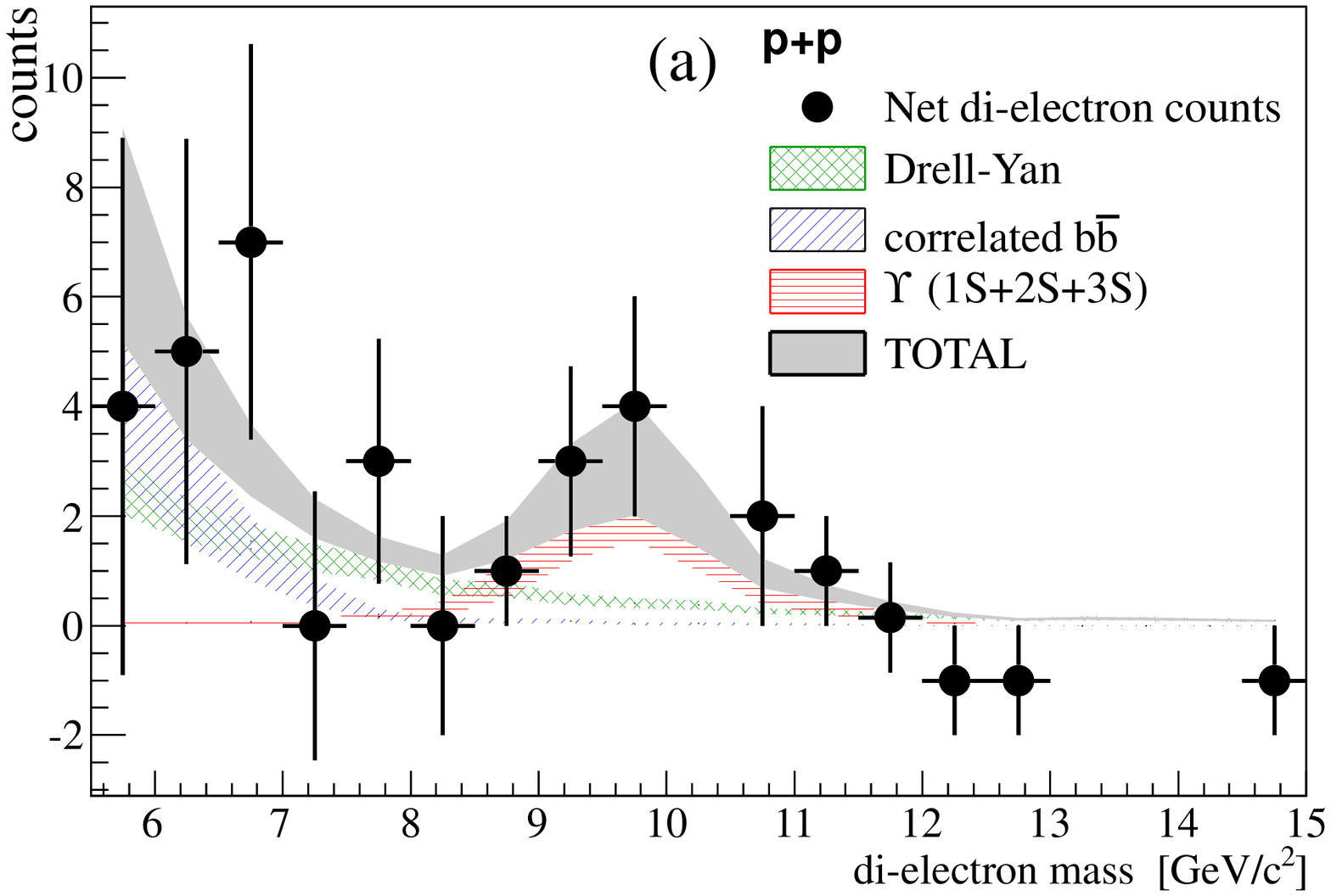}
	\includegraphics[width=0.47\linewidth]{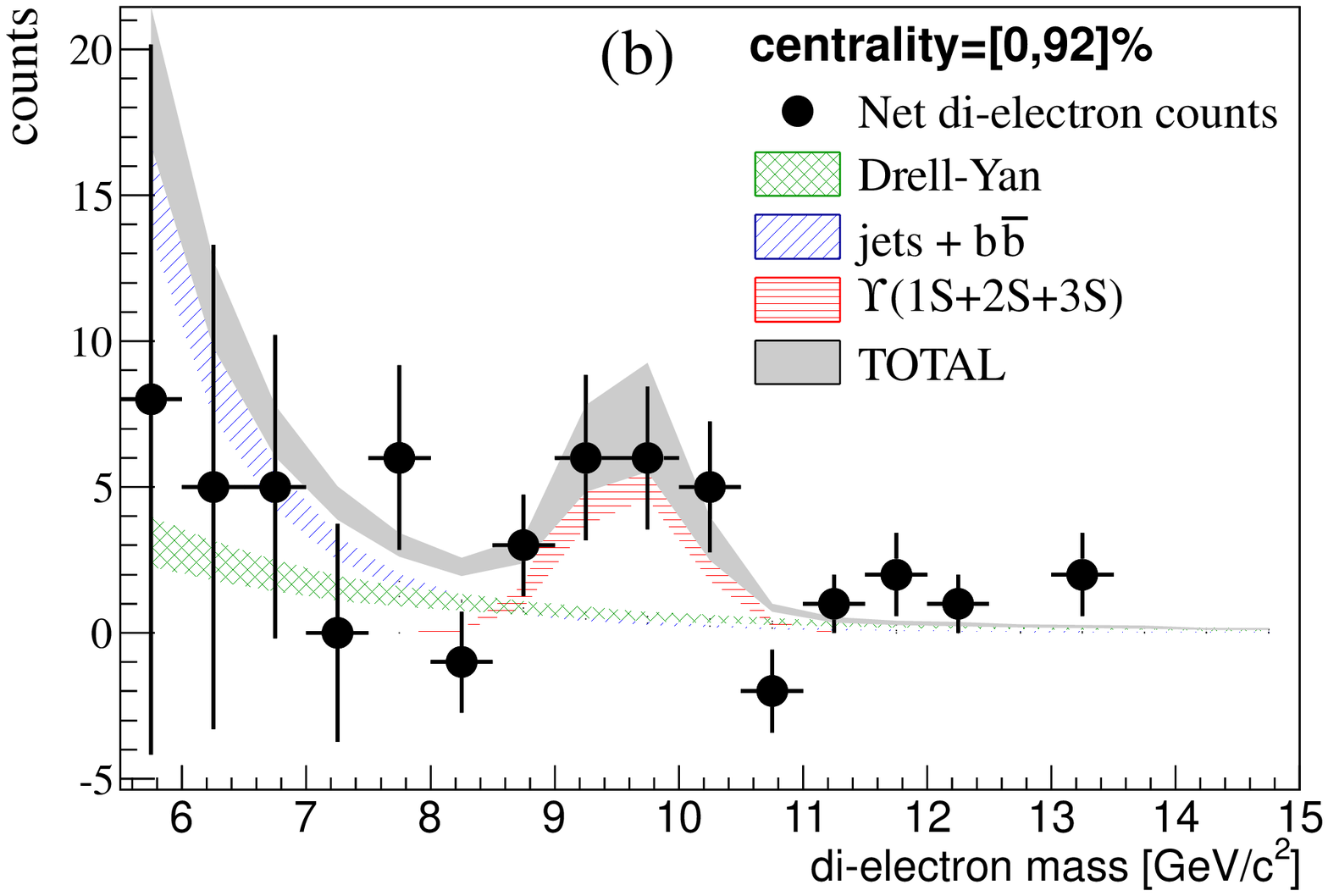}
	\includegraphics[width=0.47\linewidth]{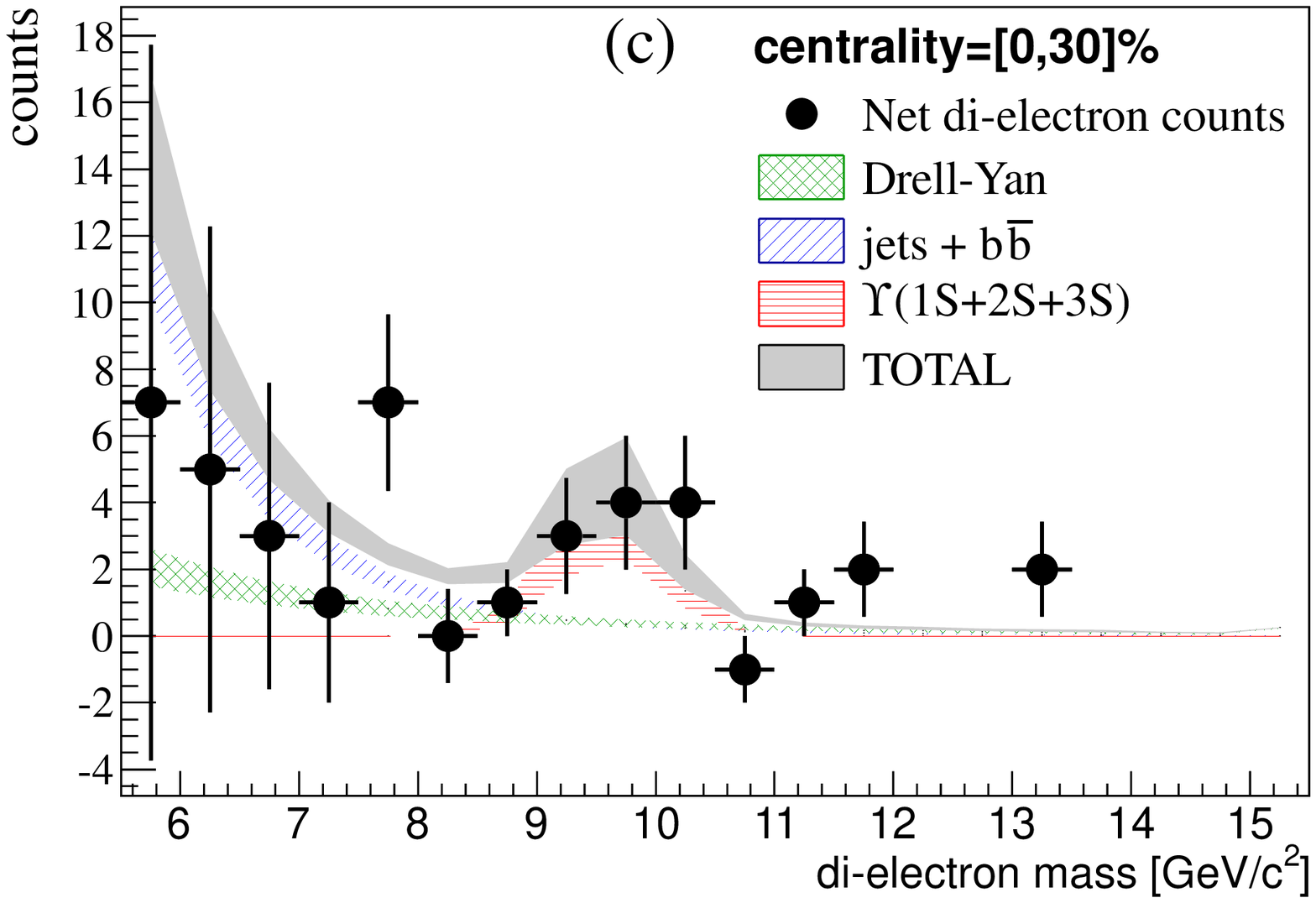}
	\includegraphics[width=0.47\linewidth]{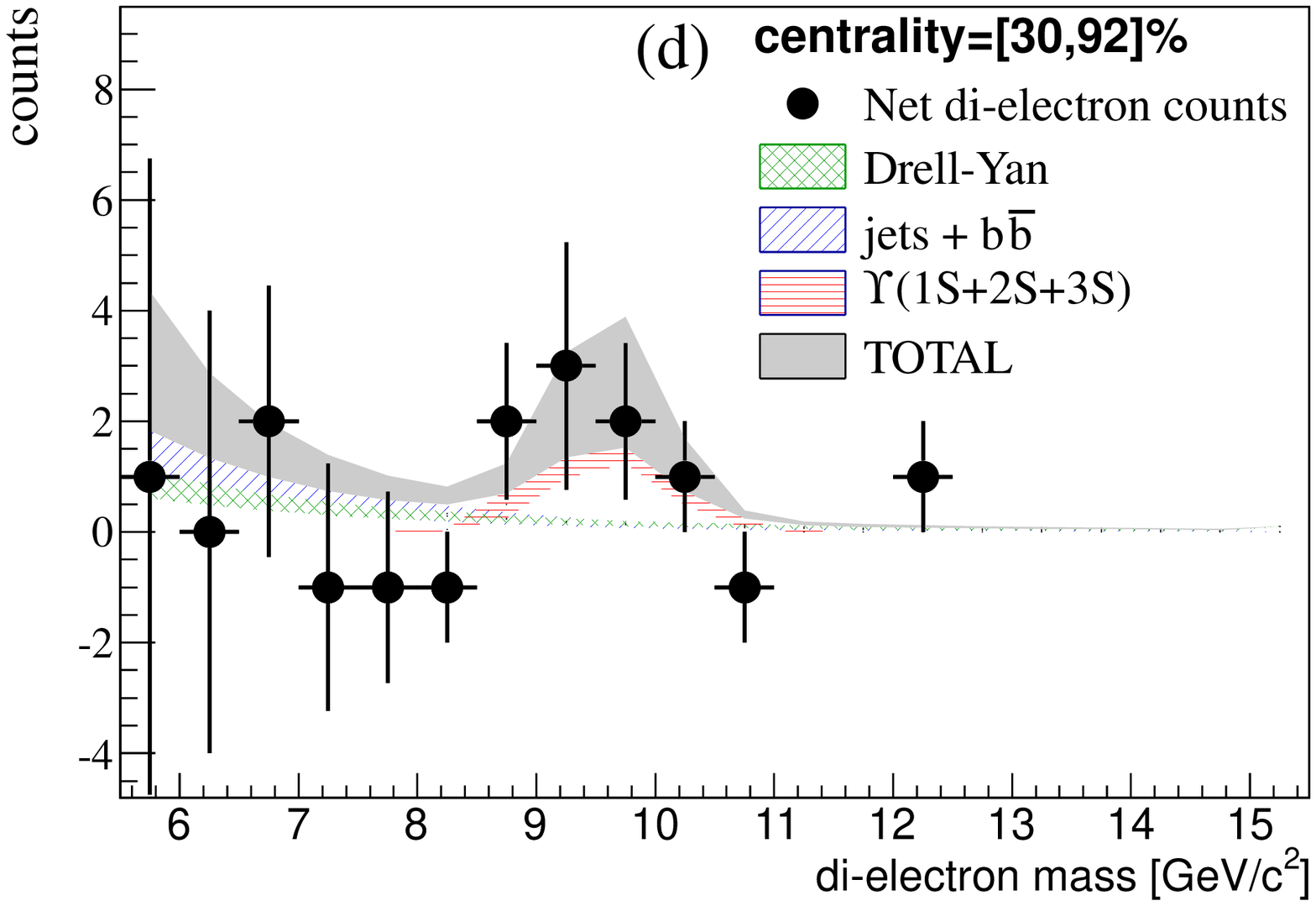}
	\caption{\label{fig:corr_fits}(Color online) Fits to the correlated
          dielectron mass distribution around the $\Upsilon$ region
          obtained in \pp collisions (a) and Au$+$Au collisions in three
          centrality bins (b,c,d). The bands correspond to fitting and
          theoretical uncertainties for the Drell-Yan estimation. Fitting
          results are used only for correlated background estimations.}
\end{figure*}
 
We cannot calculate the continuum contributions in Au$+$Au collisions in
the same way as we do for \pp collisions given the unknown nuclear
modification of bottom quarks. Contributions from correlated hadrons
may also start to be significant in a high-occupancy environment. We
thus perform a fit to separate the continuum background from the
$\Upsilon$ signal. The dielectron spectrum is described by the
following function:

\begin{eqnarray}
  \label{eq:dielectron_fit}
  f\func{m} &=&
  N_{\rm like}Y_{\rm like}\func{m} + Y_{\rm DY}(m) \\\nonumber
  &+& N_{\rm b\bar{b},jet} Y_{\rm b\bar{b},jet}(m) + Y_{\rm \Upsilon}(m)\\\nonumber
N_{\rm like} &=& \frac{2\sqrt{N_{\rm e^+e^+}N_{\rm e^-e^-}}}{\int Y_{\rm like}\func{m}
  dm}\\\nonumber
N_{\rm b\bar{b},jet} &=& \left[N_{\rm cont} -
  \int_{\rm m_{\rm low}}^{m_{\rm high}} Y_{\rm DY}(m) dm\right]\\\nonumber
Y_{\rm \Upsilon}(m) &=& \frac{N_{\rm g}}{\sqrt{2\pi}\sigma_{\rm g}} \exp\left[-\frac{1}{2}\left(\frac{m-9.5}{\sigma_{\rm g}}\right)^2\right]
\end{eqnarray}

\noindent where $N_{\rm like}\sim$1 is the normalization of the like-sign
distribution~\cite{PhysRevC81_034911}, $N_{\rm e^+e^+}+N_{\rm
  e^-e^-}=2613$ is the number of like-sign dielectron pairs over
the mass range $5<M_{\rm ee}[GeV/$c$^2]<15$ , $Y_{\rm like}(m)$ is the
like-sign dielectron mass distribution from real data which account for the
combinatorial background and a fraction of the correlated background, $Y_{\rm DY}(m)$ is the
Drell-Yan contribution as calculated in Eq. (\ref{eq:DY}),
$m_{\rm low}=8.5$ GeV/$c^2$ and $m_{\rm high}=11.5$ GeV/$c^2$ define the
mass range used in the continuum normalization, $N_{\rm cont}$ is
the continuum contribution in the $\Upsilon$ mass region,
$Y_{\rm \Upsilon}(m)$ is a Gaussian function accounting for the $\Upsilon$
peak where $\sigma_{\rm g}$ is the effective peak width of all three
$\Upsilon$ states combined, and
$Y_{\rm b\bar{b},jet}(m)$ is a function normalized in the 
$\Upsilon$ mass range which accounts for the correlated open
bottom and hadrons from jets. We assumed both a power law and an
exponential function for the correlated bottom and jet contributions:

\begin{eqnarray*}
  Y_{\rm b\bar{b},jet}(m) = \left\{\begin{array}{l}
        (\alpha+1)m^{\alpha}/\func{m_{\rm high}^{\alpha-1}-m_{\rm low}^{\alpha-1}} \\
      \alpha e^{\alpha m} / \func{e^{\alpha\cdot m_{\rm high}} - e^{\alpha
          \cdot m_{\rm low}}}\end{array}\right.
\end{eqnarray*}

The parameters $N_{\rm cont}$, $\alpha$, $N_{\rm g}$ and $\sigma_{\rm g}$ 
were fit to the unlike-sign dielectron spectrum between 5 and 16 GeV/$c^2$ 
using a maximum likelihood method. Figure~\ref{fig:corr_fits} shows the 
$f(m)-N_{\rm like}Y_{\rm like}(m)$ fitting result assuming a power law 
function for the bottom-jet contribution. The bands represent the fit and 
theoretical uncertainties. The continuum estimate changes by up to 0.9\% 
depending on the choice of the bottom+jet contribution function 
$\func{Y_{\rm b\bar{b},jet}(m)}$. Table~\ref{tab:Bdndy calc} lists the 
number of net counts and the continuum fraction for \pp and three 
centrality ranges in the Au$+$Au data. The fraction of continuum in 
Au$+$Au data obtained from these fits was found to be larger than in \pp 
data. This may reflect that the nuclear modification of Drell-Yan in 
Au$+$Au is small compared to the $\Upsilon$ yield modification.

\subsection{Mass cut efficiency}

The $\Upsilon$ count is all made in the mass range 
\mbox{$8.5<M_{\rm ee}$[GeV/$c^2$]$<11.5$}. The reconstructed $\Upsilon$ 
family peaks may have some contribution at masses out of this range. 
According to the detector simulation using the CDF 
results~\cite{PhysRevLett84_2094} for the relative yields, the mass range 
\mbox{$8.5 < M_{\rm ee}$[GeV/$c^2$]$ < 11.5$} contains a fraction 
$\varepsilon_{\rm mass}=0.94 \pm 0.05$ of the $\Upsilon(1S+2S+3S)$ yield 
in the 2006 \pp data set. The uncertainty of this estimate comes from the 
mass fit to the \pp data and from the difference between real data and 
simulations. In the Au$+$Au analysis, the evaluation of the detector 
occupancy effect on the efficiency included the mass cut used in the 
analysis. Variations in the detector mass resolution during this study 
indicate a systematic uncertainty in the mass cut efficiency of 6\% in 
Au$+$Au data. The number of $\Upsilon$ counts has a 2\% variation when the 
normalization of the like-sign dielectrons $\func{N_{\rm like}}$ is taken 
from different mass ranges. This is assigned as a systematic uncertainty 
on the yield.

\subsection{Detector Response}
\label{sec:acceptance}

The {\sc geant} based detector simulation was tuned as described
in~\cite{Adare:2011vq}. The acceptance and efficiency in this analysis
was obtained from $\Upsilon$(1S+2S+3S) dielectron decays generated by
{\sc pythia}, requiring that they fall into a rapidity range of
$|y|<$0.5. The relative yield between $\Upsilon$ states were
taken to be those reported by CDF~\cite{PhysRevLett84_2094}. This
same detector simulation was used 
to estimate the detector response for the heavy flavor and Drell-Yan
background line shapes as described in the previous section.

\begin{figure}[htb]
     \includegraphics[width=1.0\linewidth]{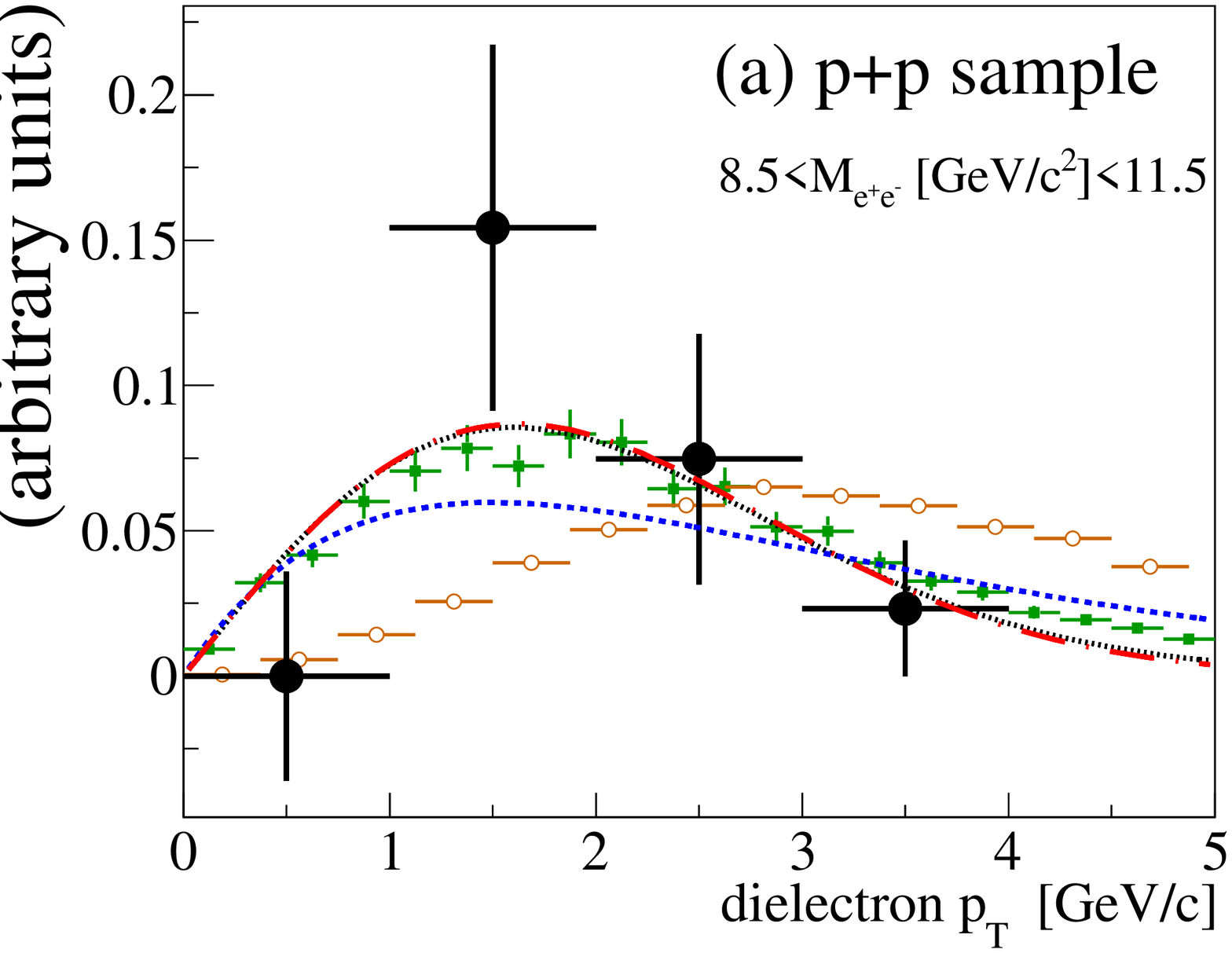}
     \includegraphics[width=1.0\linewidth]{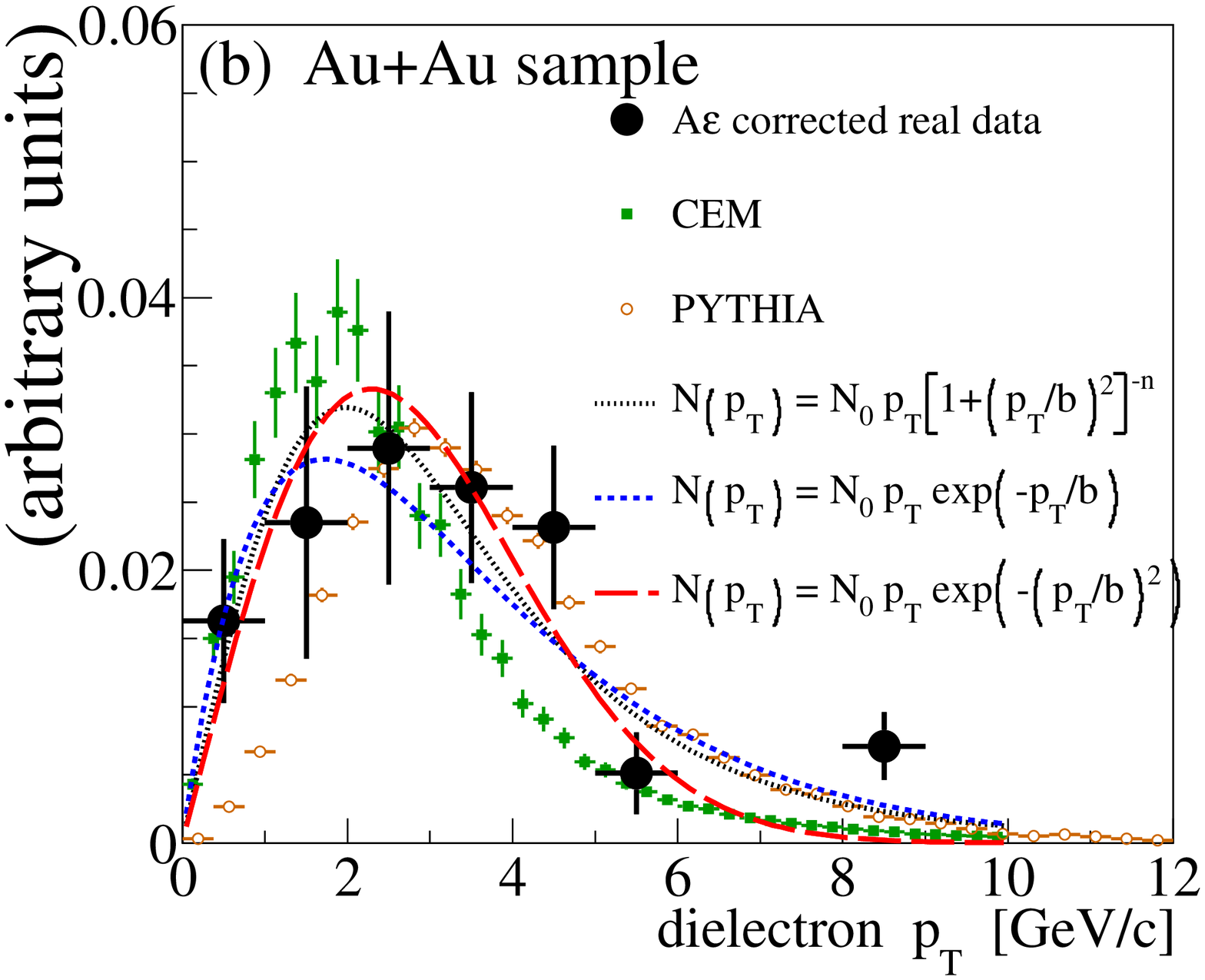}
    \caption{\label{fig:upsilon_pt} (Color online) Transverse momentum
      dependence of acceptance corrected dielectron net counts in the
      $\Upsilon$ mass region from \pp (a) and centrality integrated
      Au$+$Au (b) collisions. The lines are functions and $\Upsilon$
      yield estimations (Color Evaporation
      Model-CEM~\cite{Nelson:2012bc} and
      {\sc pythia}~\cite{Sjostrand:2006za} ) fitted to the distributions.}
\end{figure}

In the \pp sample, the overall acceptance and efficiency
$Acc\times\varepsilon$ for $\Upsilon$s calculated from simulations was
found to be (2.33 $\pm$ 0.23) \% in the $|y|<0.5$ rapidity region.  The uncertainty of this
estimate is from variations in the detector 
performance during the run, mismatches between the detector simulation
and the detector activity in real data and variations of the \pt shape
introduced in simulation (Fig.~\ref{fig:upsilon_pt}-a).

%%%%%%%%%%%%%%%%%%%%%%%%%%%%%%%%%%%%%%%%%%%%%%%%%%%%%%%%% Fig_8

The BBC trigger samples a cross section of $\sigma_{\rm pp} \times
\varepsilon_{\rm BBC} = 23 \pm 2.2$ mb in \pp collisions, according to Vernier
scans~\cite{PhysRevLett91_241803}. However, it samples a larger
fraction of the cross section when the collision includes a hard
scattering process. Studies with high \pt $\pi^0$ yields showed an
increase of the luminosity scanned by the BBC by a factor of
\mbox{$1/\varepsilon_{\rm BBC^{hard}}$, $\varepsilon_{\rm BBC}^{hard}
  = (0.79 \pm 0.02)$}~\cite{Adare:2010de}. In Au$+$Au data the BBC scans
$92 \pm 3$\% of the total Au$+$Au inelastic cross section and there is
no bias from hard scattering ($\varepsilon_{\rm BBC}^{hard}$=1). 
The EMCal-RICH trigger (ERT) efficiency of dielectrons was
found to be (79.6 $\pm$ 3.6)\% in the \pp sample when emulating the
ERT in MB data. The ERT was not used for the Au$+$Au data.

%%%%%%%%%%%%%%%%%%%%%%%%%%%%%%%%%%%%%%%%%%%%%%%%%%%%%%%%% Fig_9
\begin{figure}[tbh]
     \includegraphics[width=1.0\linewidth]{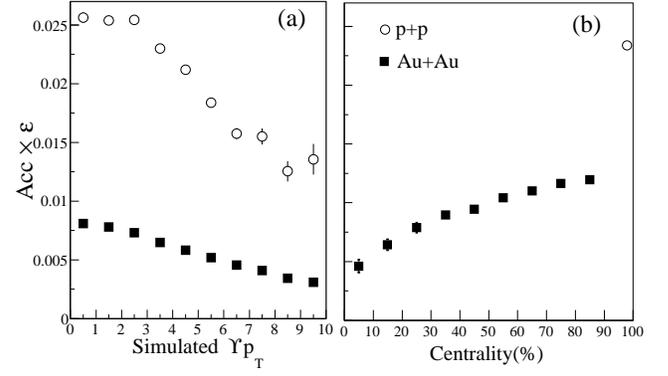}
    \caption{\label{fig:eff} Dependence of the acceptance $\times$
      efficiency for detected $\Upsilon$ dielectron decays in \pp and
      Au$+$Au collisions on (a) transverse momentum in 0\%--92\% centrality
      and (b) collision centrality. The bars represent statistical
      uncertainties in the simulation.}
\end{figure}

In the Au$+$Au data, the electron identification cuts were tighter,
resulting in a calculated acceptance and efficiency
$Acc\times\varepsilon=1.41 \pm 0.05$\% (point at 85\% centrality in
Fig.~\ref{fig:eff}-b). To quantify additional
inefficiencies from the detector occupancy, the raw detector signal
from simulated $\Upsilon$ dielectron decays was embedded in real raw
data. The simulated $\Upsilon$ was generated at the same collision
point measured in the real event. The reconstruction, fitting and mass
cuts of the embedded data were the same as those used in real data
analysis.  The $p_T$ and collision centrality dependence of the
resulting fraction of $\Upsilon$ counts in the reconstructed embedded
data are shown in Fig.~\ref{fig:eff}. The big difference between the
detector efficiency obtained in \pp data and peripheral Au$+$Au reflects
the tight cuts needed in Au$+$Au because of the larger occupancy and
additional material in front of the detector in 2010 run.
 
Because we do not have the statistic precision to determine the
transverse momentum distribution of the $\Upsilon$, we must
employ models for the $p_T$ dependence to determine an
overall acceptance and efficiency. Five functions were used for the
$p_T$ distribution: a shape from generated $\Upsilon$ decays in
{\sc pythia}, a prediction from the color evaporation model~\cite{Nelson:2012bc}
and three fitted functions $f\func{p_T}$ to the acceptance corrected real data
distribution (Fig.~\ref{fig:upsilon_pt}). The $p_T$ integrated acceptance and
efficiency is determined by an average using the $p_T$ dependence
shown in Fig.~\ref{fig:eff} and these functions as
weights. The difference between these calculations and the
default weighing using {\sc pythia} as an input is within 7.8\% in \pp and 7.9\%
in Au$+$Au samples. 

The final values for the efficiency in our wide centrality bins are also 
sensitive to the true centrality dependence of the $\Upsilon$ production.  
To estimate this systematic uncertainty we assume two different centrality 
dependence models: (1) binary collision scaling and (2) participant 
collision scaling. Within our centrality ranges, we find that these two 
models yield less than a 7\% difference and we include this in our 
occupancy systematic uncertainty.

\section{\label{sec:results}Results}

The $\Upsilon\rightarrow e^+e^-$ invariant multiplicity at midrapidity,
BdN/dy, is calculated by

\begin{equation}
  B\frac{dN}{dy} = \frac{1}{\Delta y}
  \frac{N_{\rm \Upsilon}}{(N_{\rm BBC}/c) \cdot Acc\cdot \varepsilon}
  \label{eq:bdndy}
\end{equation}

\noindent where $B$ is the dielectron branching ratio, $N_{\rm \Upsilon}$  is the number of $\Upsilon$ candidates
in the data set as defined in (\ref{eq:upsilon_counts}), $\Delta y=1$
corresponds to the rapidity range used in simulation ($\pm$0.5),
$N_{\rm BBC}$ is the number of analyzed events, 
$c=\varepsilon_{\rm BBC}/\varepsilon_{\rm BBC}^{\rm hard}$ is a
correction factor accounting for the limited BBC efficiency and the
trigger bias present in events which contain a hard scattering in \pp
collisions as explained in Section \ref{sec:acceptance}, $Acc$
is the $\Upsilon$ acceptance and $\epsilon$ is
the $\Upsilon$ reconstruction efficiency which includes the ERT efficiency. Table~\ref{tab:Bdndy calc}
summarizes the numbers used to calculate the $\Upsilon$ yields 
using Eq.~\ref{eq:bdndy}. Table~\ref{tab:sys_error} details the systematic
uncertainties involved in the yield calculation.
The resulting invariant
multiplicities are reported in Table~\ref{tab:Bdndyresults}.  

%%%%%%%%%%%%%%%%%%%%%%%%%%%%%%%%%%%%%%%%%%%%%%%%%%%%% Table_III
\begin{table*} [t]
   \caption{\label{tab:Bdndy calc} Summary of values used in BdN/dy
    (\ref{eq:bdndy}) and $R_{AA}$ (\ref{eq:raa}) calculations.} 
    \begin{ruledtabular} \begin{tabular}{ c  c  c  c  c }
    Value &  \pp & Au$+$Au 0\%--92\% & Au$+$Au 0\%--30\% & Au$+$Au 30\%--92\% \\
    \hline
    $N_{\rm unlike}-N_{\rm like}$ & 10.5$^{+3.7}_{-3.6}$ & 18.3$^{+5.0}_{-5.2}$ & 11.2$^{+3.8}_{-4.0}$ & 6.4$^{+3.3}_{-3.5}$ \\
    $f_{\rm cont}$                  & 0.13 $\pm$ 0.04     & 0.216 $\pm$
    0.045          & 0.270 $\pm$ 0.063      & 0.186 $^{+0.065}_{-0.060}$   \\
    $N_{\rm BBC} \times 10^9$ & 143  & 5.40       & 1.62     & 3.35 \\
    $c$&0.70&1&1&1\\
    $Acc\times\varepsilon$     &    (1.64 $\pm$ 0.25)\%   & (0.65 $\pm$ 0.13)\%
    & (0.58 $\pm$ 0.11)\%   &    (0.96 $\pm$ 0.18)\%    \\
    \Ncoll & 1 & 258 $\pm$ 25 & 644 $\pm$ 63 & 72 $\pm$ 7 \\
    \Npart& 2& 109 $\pm$ 4 & 242 $\pm$ 4 & 45 $\pm$ 2\\
    \end{tabular} \end{ruledtabular}
\end{table*}

%%%%%%%%%%%%%%%%%%%%%%%%%%%%%%%%%%%%%%%%%%%%%%%%%%%%% Table_IV
\begin{table} [htb]
   \caption{\label{tab:sys_error} Summary of the relative systematic
     uncertainties involved in $BdN/dy$ calculations.} 
    \begin{ruledtabular} \begin{tabular}{lccc}
      & \multicolumn{2}{c}{Uncertainty} \\
      Systematic & \pp & Au$+$Au \\ \hline
      acceptance & 7.5\% & 7.0\% \\
      electron identification & 1.1\% & 5.0\% \\
      simulation input & 7.8\% & 7.9\% \\
      mass cut efficiency & 6.3\% & 5.0\% \\
      continuum contribution & 5\% & 5.8\%--8.6\%\\
      acceptance fluctuation & 7.3\% & 14.0\% \\
      ERT efficiency & 4.5\% & NA \\
      occupancy effect & NA & 2.0\%--7.5\%\\
      combinatorial background & 2.0\% & 2.0\% \\
\\
      TOTAL & 16.1\% & 20.7-21.2\% \\ 
    \end{tabular} \end{ruledtabular}
\end{table}

The $\Upsilon$(1S+2S+3S) cross section in \pp collisions is 

\begin{eqnarray}
  \left. B\frac{d\sigma_{\rm \Upsilon}}{dy}\right|_{\rm |y|<0.5} &=& B \frac{dN}{dy} \times
  \sigma_{\rm pp} \\\nonumber
  &=& 108  \pm 38 \textrm{(stat)} \pm 15 \textrm{(syst)} \pm 11 \textrm{(lum) pb},
\end{eqnarray}

\noindent where $\sigma_{\rm pp}$= 42mb is the \pp inelastic cross section
at \fullb.

\begin{figure}
  \includegraphics[width=1.0\linewidth]{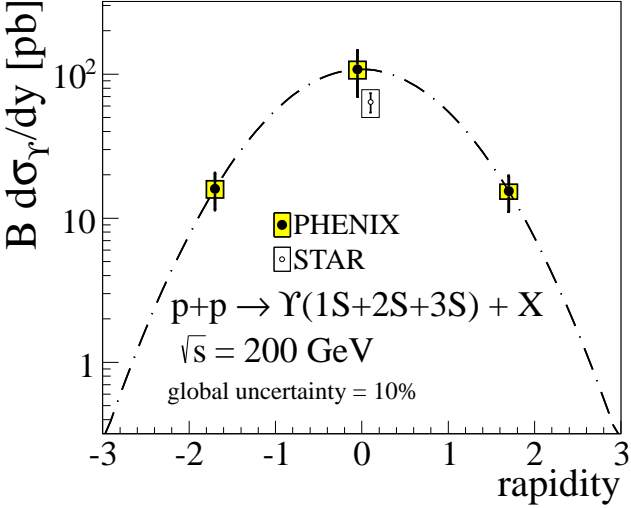}
  \caption{\label{fig:upsilon_results} (Color online) Rapidity dependence of
    $\Upsilon$(1S+2S+3S) yield measured by PHENIX, forward rapidity
    result from~\cite{PhysRevC87_044909} and STAR midrapidity
    from~\cite{tagkey2014127}. 
    Dashed line is a Gaussian function fitted to the points. The points 
    at zero rapidity are shifted for clarity.}
\end{figure}

\begin{figure}
  \includegraphics[width=1.0\linewidth]{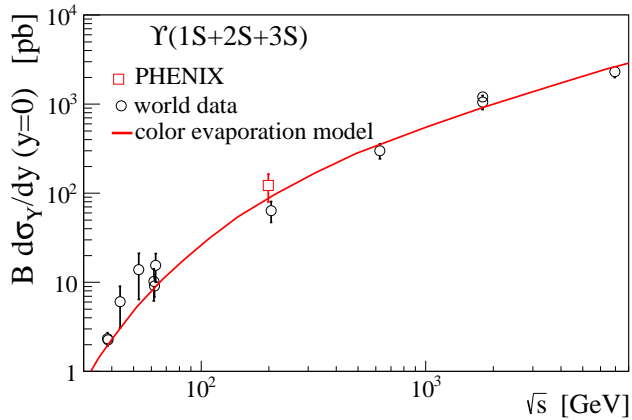}
  \caption{\label{fig:upsilon_sqrt} (Color online) Energy dependence
    of the $\Upsilon$(1S+2S+3S) differential cross section at
    midrapidity in \pp and $p$+$\bar{p}$
    collisions~\cite{PhysRevD43_2815,PhysRevLett41_684,PhysRevLett42_486,PhysRevLett55_1962,PhysRevD39_3516,Kourkoumelis:1980hg,Angelis:1979ar,Albajar:1986iu,Acosta:2001gv,tagkey2014127,Khachatryan:2010zg}. The
    curve is the estimation using the color evaporation
    model~\cite{Nelson:2012bc}.}
\end{figure}

Figure~\ref{fig:upsilon_results} shows the rapidity dependence of 
$\Upsilon$ measured in \pp collisions by PHENIX in the mid- (this 
analysis), forward rapidities~\cite{PhysRevC87_044909} and the STAR
result at midrapidity \cite{tagkey2014127}.  
Figure~\ref{fig:upsilon_sqrt} presents the collision energy dependence of 
the differential cross section at midrapidity along with a NLO calculation 
using the color evaporation model for the bottomonium 
hadronization~\cite{Nelson:2012bc}.

%%%%%%%%%%%%%%%%%%%%%%%%%%%%%%%%%%%%%%%%%%%%%%%%%%%%%%%%% Fig_10

%%%%%%%%%%%%%%%%%%%%%%%%%%%%%%%%%%%%%%%%%%%%%%%%%%%%%%%%% Fig_11

%%%%%%%%%%%%%%%%%%%%%%%%%%%%%%%%%%%%%%%%%%%%%%%%%%%%% Table_V
\begin{table}
 \caption{Summary of the measured $\Upsilon$ invariant
   multiplicities, $B dN/dy$, for one \pp three Au$+$Au data sets.\label{tab:Bdndyresults}} 
    \begin{ruledtabular} \begin{tabular}{ c  c  }
    Centrality & $B dN/dy$ \\
    \hline
     \pp $\func{\times 10^{9}}$ & 2.7 $\pm$ 0.9 (stat)  $\pm$ 0.4 (syst)\\
    0\%--92\% $\func{\times 10^{7}}$ & 4.1$^{+1.1}_{-1.2}$ (stat) $\pm$ 0.9  (syst)\\
    0\%--30\% $\func{\times 10^{7}}$ & 8.7 $^{+2.9}_{-3.1}$ (stat)$\pm$ 1.8 (syst)\\
    30\%--92\% $\func{\times 10^{7}}$ & 1.6 $^{+0.8}_{-0.9}$ (stat)
    $\pm$ 0.3 (syst)\\
    \end{tabular} \end{ruledtabular}
 \end{table}

\begin{figure}[tbh]
    \includegraphics[width=1.0\linewidth]{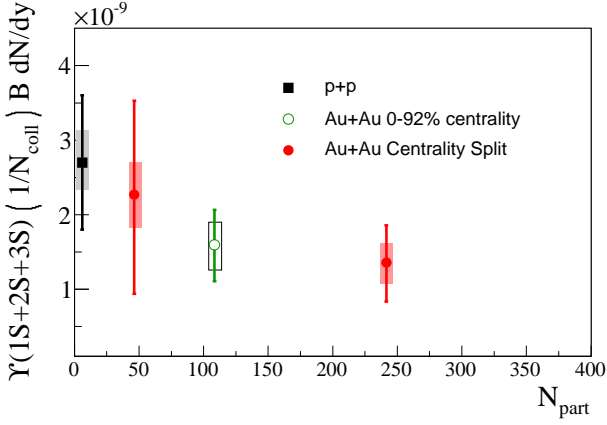}
    \caption{(Color online) The N$_{\rm coll}$ normalized invariant yield of $\Upsilon$s
      produced during the 2006 \pp and the 2010 Au$+$Au operations, as
      a function of N$_{\rm part}$.\label{fig:bdndy}. } 
\end{figure}

In addition to the Au$+$Au  0\%--92\% centrality sample, 
we present data in two centrality bins, 0\%--30\% most central and
30\%--92\% most central. 
Using a Monte Carlo simulation based on the Glauber model
in~\cite{Miller:2007ri}, we estimated \Ncoll, the average number of
binary nucleon-nucleon collisions and \Npart, the average number of
participants, for all data samples.  Figure~\ref{fig:bdndy} shows the
\Ncoll normalized invariant yield of $\Upsilon$ decays as a function
of the number of participants. For central Au$+$Au collisions, we
observe a reduction of the yield relative to a pure \Ncoll
scaling that is typical of hard scattering processes.

%%%%%%%%%%%%%%%%%%%%%%%%%%%%%%%%%%%%%%%%%%%%%%%%%%%%%%%%% Fig_12

The nuclear modification factors for the binned and
integrated 0\%--92\% centrality data set $\func{R_{AA}}$  were calculated as: 

\begin{equation}
  R_{AA} = \frac{dN/dy_{\rm AuAu}}{<N_{\rm coll}> dN/dy_{\rm pp}}
  \label{eq:raa}
\end{equation}
and are reported in Table~\ref{tab:raaresults}. A global uncertainty
of 40\% is obtained from the quadratic sum of the relative uncertainty from 38\% \pp data
(statistical+systematic) and 12\% from the Glauber estimate of the number of
collisions. We assume none of the systematic uncertainties are correlated
between \pp and Au$+$Au samples given the different collision
environment and changes in the detector configuration between 2006 and
2010 runs, namely active area differences and the installation of the
hadron blind detector in 2010 which increased the radiation length
from 0.4\% to 2.8\%.

%%%%%%%%%%%%%%%%%%%%%%%%%%%%%%%%%%%%%%%%%%%%%%%%%%%%% Table_VI
\begin{table}
  \caption{Summary of the measured $\Upsilon$ nuclear modification
    factors, $R_{AA}$, for Au$+$Au data sets.  \label{tab:raaresults}}
    \begin{ruledtabular} \begin{tabular}{ c  c  }
    Centrality & $R_{AA}$ \\
    \hline
    0\%--92\% &0.58$\pm$ 0.17(stat) $\pm$ 0.13 (syst) $\pm$ 0.23 (global)\\
    0\%--30\%& 0.50$\pm$ 0.18 (stat)$\pm$ 0.11 (syst) $\pm$ 0.20 (global)\\
    30\%--92\%& 0.84  $^{+0.45}_{-0.48}$ (stat) $\pm$ 0.18 (syst)
    $\pm$ 0.34 (global)\\
    \end{tabular} \end{ruledtabular}
\end{table}

\begin{figure}[tbh]
    \includegraphics[width=1.0\linewidth]{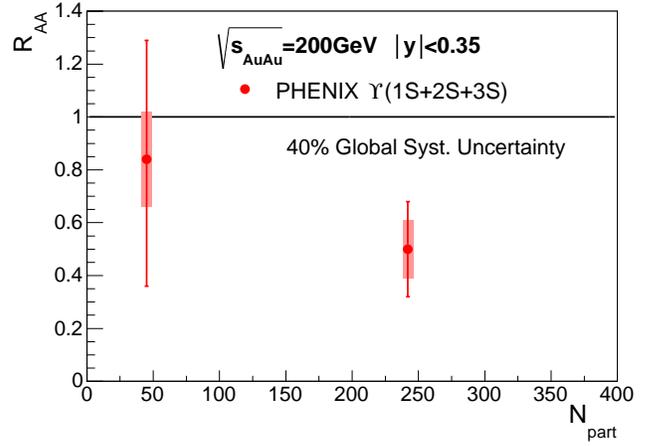}
  \caption{(Color online) Nuclear modification factor for centrality binned data
    plotted as a function of N$_{\rm part}$.\label{fig:centdepraa}} 
\end{figure}

If the $\Upsilon(1S+2S+3S)$ yield for Au$+$Au collisions is equal to the
yield for \pp collisions times the number of
binary collisions in Au$+$Au collisions, then \raa=1 and there are no
nuclear modification effects. Figure~\ref{fig:centdepraa} shows the
\raa as a function of the number of participants for the two
centrality-split classes. The inclusive
$\Upsilon$ states are suppressed in central 200 GeV Au$+$Au
collisions, corresponding to large $N_{\rm part}$. However, the degree of
suppression in semi-peripheral collisions is unclear, due to limited
statistics.

%%%%%%%%%%%%%%%%%%%%%%%%%%%%%%%%%%%%%%%%%%%%%%%%%%%%%%%%% Fig_13

In most central events, the suppression is comparable to what is
observed in $p(d)$+A
collisions~\cite{PhysRevD43_2815,PhysRevLett66_2285,PhysRevLett100_062301,PhysRevC87_044909}. 
Based on the lattice calculations discussed before,
the bottomonia excited states should be completely dissociated in the
core of Au$+$Au collisions at RHIC. Table~\ref{tab:raa_assumption}
summarizes what would be the $R_{AA}$ observed in this study in
case the only nuclear matter effect observed is the complete
suppression of these excited states. The estimation is based on the
composition of the $\Upsilon$ states measured and the decays to the
$\Upsilon$(1S) reported in Tables~\ref{tab:upsilon_frac} and
\ref{tab:upsilon1s_source}. The $R_{AA}$ obtained in this analysis
is consistent with the suppression of excited states if other
initial and final state effects are ignored.

%%%%%%%%%%%%%%%%%%%%%%%%%%%%%%%%%%%%%%%%%%%%%%%%%%%%% Table_VII
\begin{table}
  \caption{\label{tab:raa_assumption} $\Upsilon$(1S+2S+3S) $R_{AA}$
    expected when the excited states are completed suppressed in Au$+$Au
  collisions along with the measured result in the 30\% most central
  collision regime. Estimations based on Tables~\ref{tab:upsilon_frac} and
  \ref{tab:upsilon1s_source}.}
  \begin{ruledtabular} \begin{tabular}{lc}
    & $R_{AA}$ \\\hline
    no 2S or 3S & 0.65 $\pm$ 0.11\\
    no 2S,3S or $\chi_B$ & 0.37 $\pm$ 0.09\\
    measured & 0.50$\pm$0.18 (stat)$\pm$ 0.11 (syst) $\pm$ 0.19(global)\\
  \end{tabular} \end{ruledtabular}
\end{table}

\begin{figure}[tbh]
    \includegraphics[width=1.0\linewidth]{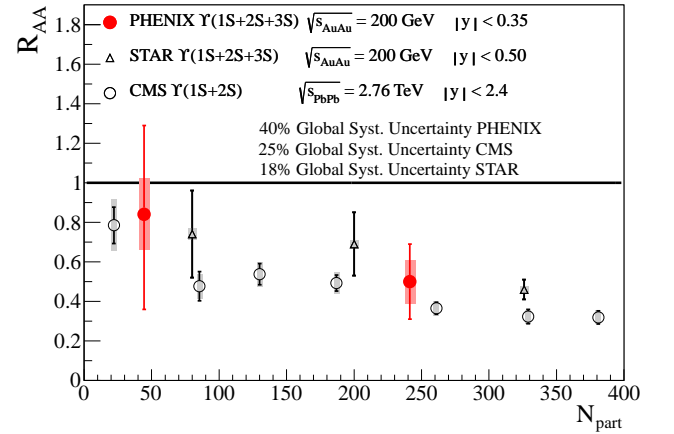}
  \caption{(Color online) 
Nuclear modification factor for centrality binned data plotted as a 
function of N$_{\rm part}$ compared to STAR 
\cite{tagkey2014127} and CMS \label{fig:cms-comp-raa} results.}
\end{figure}

The result presented in this work agrees
with the STAR experiment at the same energy \cite{tagkey2014127}.
The CMS experiment reported centrality dependent nuclear modification 
factors for the separated $\Upsilon$(1S) and $\Upsilon$(2S) states at 
\mbox{$\sqrt{s_{_{NN}}}$=2.76 TeV} in Pb$+$Pb collisions at the 
LHC~\cite{Chatrchyan:2012lxa}.  CMS also reported an 
upper limit of $R_{AA}(\Upsilon(3S))$ of 0.10 at the 95\% confidence 
level.  Figure~\ref{fig:cms-comp-raa} compares the observed inclusive 
$\Upsilon$(1S+2S+3S) nuclear modification factor observed by PHENIX with 
STAR and the inclusive $\Upsilon$(1S+2S) measurement by CMS at higher 
energy showing that the observed nuclear modification factors are very 
similar at the two quite different energies.

%%%%%%%%%%%%%%%%%%%%%%%%%%%%%%%%%%%%%%%%%%%%%%%%%%%%%%%%% Fig_14

\begin{figure}[tbh]
	\includegraphics[width=1.0\linewidth]{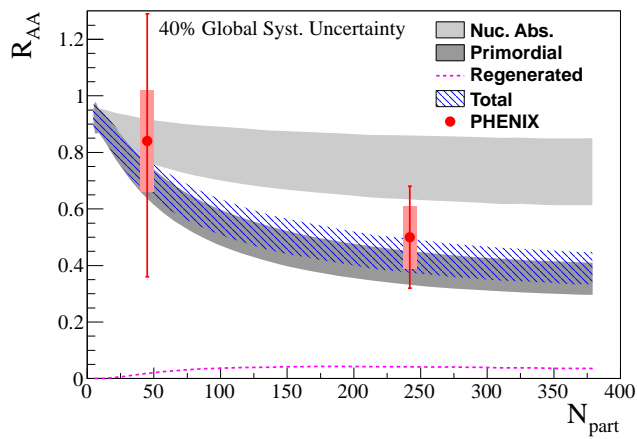}
	\caption{(Color online) A comparison of PHENIX data to the
          model from~\cite{Emerick:2011xu} for the strong binding
          scenario\label{fig:rapp_sbs_comp}.}
\end{figure}

\begin{figure}[tbh]
	\includegraphics[width=1.0\linewidth]{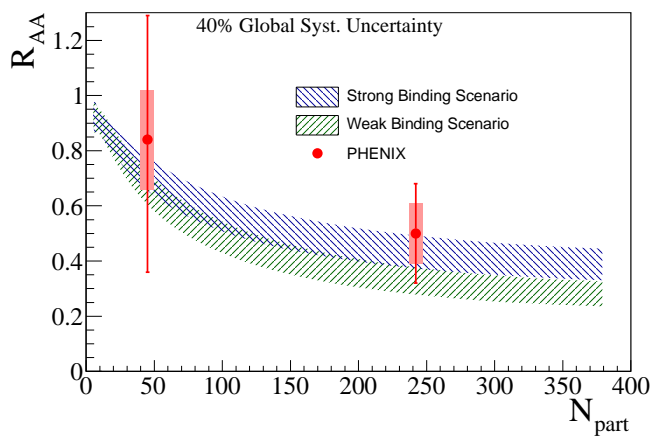}
	\caption{(Color online) A comparison of PHENIX $\Upsilon$ data to the model
          from~\cite{Emerick:2011xu} for the weak and strong binding
          scenario\label{fig:rapp_wbs_comp}.} 
\end{figure}

Additionally, it is important to compare the measurements to various model 
predictions. A model by R. Rapp {\it et al.} has frequently been used to 
interpret J/$\psi$ production~\cite{Emerick:2011xu}.  It uses a 
rate-equation approach, which accounts for both suppression from cold nuclear 
matter, color screening of excited states (seen in 
Fig.~\ref{fig:qqbar_diccociation_T}) and regeneration mechanisms in the 
QGP and hadronization phases of the evolving medium. This study looked at 
two scenarios. The first is the strong binding scenario where the 
bottomonium binding energy was not affected by the presence of the QGP, 
remaining at the values found in vacuum, and is shown in 
Fig.~\ref{fig:rapp_sbs_comp}. The other is the weak binding scenario where 
the bottomonium bound-state energies are significantly reduced in the QGP, 
relative to the vacuum state, adopting the screened Cornell-potential 
results of~\cite{Karsch:1987pv} and is shown in 
Fig.~\ref{fig:rapp_wbs_comp}.  Our data, albeit with large statistical 
uncertainties, are consistent with both versions of this model.

%%%%%%%%%%%%%%%%%%%%%%%%%%%%%%%%%%%%%%%%%%%%%%%%%%%%%%%%% Fig_15

%%%%%%%%%%%%%%%%%%%%%%%%%%%%%%%%%%%%%%%%%%%%%%%%%%%%%%%%% Fig_16

\begin{figure}
 		\includegraphics[width=1.0\linewidth]{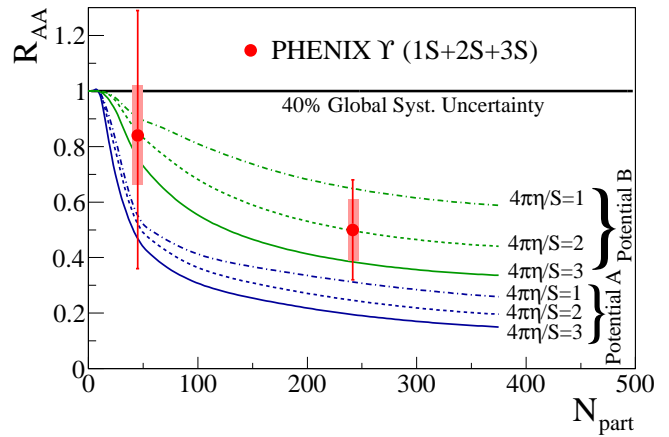}
 		\caption{(Color online) Centrality dependent $R_{AA}$ compared to
                  model predictions from Strickland and
                  Bazow~\cite{Strickland:2011aa} \label{fig:theorycomparison}.} 
 \end{figure}

More recently, two new models were suggested by Strickland and
Bazow~\cite{Strickland:2011aa} based on the potential
model~\cite{Karsch:1987pv}, with the addition of an anisotropic
momentum term. Models A and B are identical, except for an additional
term in Model B which adds an entropy contribution to the free
energy.  Figure~\ref{fig:theorycomparison} shows the PHENIX measurement
along with the two model predictions, each with a variety of values
for the ratio of the shear viscosity to the entropy density. No
definitive statement can be made regarding the shear
viscosity. However, the extreme potential B case appears to be favored.

%%%%%%%%%%%%%%%%%%%%%%%%%%%%%%%%%%%%%%%%%%%%%%%%%%%%%%%%% Fig_17

\section{Conclusions}

In summary, we have studied the production of the sum of $\Upsilon$ states
1S, 2S and 3S at \full in the midrapidity region. The dielectron
channel differential cross section in \pp collisions is $B d\sigma/dy=$108 $\pm$ 38 (stat) $\pm$ 15 (syst) $\pm$ 11
(luminosity) pb. The nuclear modification seen in Au$+$Au
minimum bias collisions is 0.58 $\pm$ 0.17 (stat) $\pm$ 0.13(syst) $\pm$ 0.23
(global), whereas it is 0.84 $^{+0.45}_{-0.48}$ (stat) $\pm$ 0.18
(syst) $\pm$ 0.34(global) in the mid-peripheral events and
0.50 $\pm$ 0.18 (stat)$\pm$ 0.11 (syst) $\pm$ 0.20(global) in
the 30\% most central events. The nuclear modification is consistent with the
complete suppression of the bottomonium excited states
($\Upsilon$(2S), $\Upsilon$(3S) and $\chi_B$), in qualitative
agreement with most calculations as compiled in
Fig.~\ref{fig:qqbar_diccociation_T}, assuming no cold nuclear
matter effects. There are several detailed model calculations that show good
agreement with our measured modifications.  The nuclear modification factors
measured by PHENIX are similar to measurements by STAR at the same energy 
and by CMS at much higher energy, $\sqrt{s_{\rm NN}}$=2.76 TeV.

%%%%%%%%%%%%%%%%%%%%%%%%%  Acknowledgements 

\section*{ACKNOWLEDGMENTS} % Run-6,Run-10 long form for all journals

We thank the staff of the Collider-Accelerator and Physics
Departments at Brookhaven National Laboratory and the staff of
the other PHENIX participating institutions for their vital
contributions.  We acknowledge support from the 
Office of Nuclear Physics in the
Office of Science of the Department of Energy,
the National Science Foundation, 
a sponsored research grant from Renaissance Technologies LLC, 
Abilene Christian University Research Council, 
Research Foundation of SUNY, and
Dean of the College of Arts and Sciences, Vanderbilt University 
(U.S.A),
Ministry of Education, Culture, Sports, Science, and Technology
and the Japan Society for the Promotion of Science (Japan),
Conselho Nacional de Desenvolvimento Cient\'{\i}fico e
Tecnol{\'o}gico and Funda\c c{\~a}o de Amparo {\`a} Pesquisa do
Estado de S{\~a}o Paulo (Brazil),
Natural Science Foundation of China (P.~R.~China),
Croatian Science Foundation and
Ministry of Science, Education, and Sports (Croatia),
Ministry of Education, Youth and Sports (Czech Republic),
Centre National de la Recherche Scientifique, Commissariat
{\`a} l'{\'E}nergie Atomique, and Institut National de Physique
Nucl{\'e}aire et de Physique des Particules (France),
Bundesministerium f\"ur Bildung und Forschung, Deutscher
Akademischer Austausch Dienst, and Alexander von Humboldt Stiftung (Germany),
Hungarian National Science Fund, OTKA, and 
the Hungarian American Enterprise Scholarship Fund (Hungary), 
Department of Atomic Energy and Department of Science and Technology (India),
Israel Science Foundation (Israel), 
National Research Foundation and WCU program of the 
Ministry Education Science and Technology (Korea),
Physics Department, Lahore University of Management Sciences (Pakistan),
Ministry of Education and Science, Russian Academy of Sciences,
Federal Agency of Atomic Energy (Russia),
VR and Wallenberg Foundation (Sweden), 
the U.S. Civilian Research and Development Foundation for the
Independent States of the Former Soviet Union, 
the US-Hungarian Fulbright Foundation for Educational Exchange,
and the US-Israel Binational Science Foundation.

%%%%%%%%%%%%%%%%%%%%%%%%%%%  References 

%\bibliography{ppg111x3}   

 
\end{document}